\documentclass[twoside,twocolumn,english,superscriptaddres,showpacs,longbibliography,aps,prb]{revtex4-2}

\usepackage[T1]{fontenc}
\usepackage[utf8]{inputenc}
\usepackage[english]{babel}
\usepackage{bm}
\usepackage{amsmath}
\usepackage{amssymb}
\usepackage{graphicx}
\usepackage[unicode=true, pdfusetitle, bookmarks=true,bookmarksnumbered=true,bookmarksopen=true,bookmarksopenlevel=1,
breaklinks=false,pdfborder={0 0 0},pdfborderstyle={},backref=false,color links, citecolor=blue,linkcolor=red,urlcolor=blue] {hyperref}
\usepackage[nameinlink]{cleveref}
\Crefname{figure}{Fig.}{}

\makeatletter
\def\set@firstnote#1{%
 \@ifnum{\firstnote@num=#1\relax}{}{%
  \class@warn@end{Endnote numbers changed: rerun LaTeX}%
 }%
 \immediate\write\@mainaux{%
   \global\mathchardef\string\firstnote@num#1\relax
 }%
}%

\usepackage{braket}
\usepackage{bm}
\usepackage{tikz}
\usepackage{pgfplots}
\usepackage{pgf}
\usepackage{subfigure}
\usepackage{nicematrix}
\usepackage{diagbox}
\usepackage{orcidlink}

\renewcommand{\selectlanguage}[1]{}

\makeatother

\pgfplotsset{compat=1.18}
\begin{document}

\title{Gauging or extending bulk and boundary conformal field theories: Application to bulk and domain wall problem in topological matter and their descriptions by (mock) modular covariant}
\author{Yoshiki Fukusumi}
\affiliation{Physics Division, National Center of Theoretical Sciences, National Taiwan University, Taipei 106319, Taiwan}
\affiliation{Center for Theory and Computation, National Tsing Hua University, Hsinchu 300044, Taiwan}

\date{\today}
\begin{abstract}
We formulate the $Z_{N}$ group gauging operations (or group extensions) in (smeared) boundary conformal field theories (BCFTs) and bulk conformal field theories, and study their applications to various phenomena in topologically ordered systems. We apply the resultant theories to analyze the correspondence between the renormalization group (RG) flow of CFTs and the classification of topological quantum field theories with the testable information of general classes of partition functions. One can obtain the bulk topological properties of $2+1$ dimensional topological ordered phase corresponding to the massive RG flow of $1+1$ dimensional systems equivalent to the smeared BCFT. We present an obstruction to mass condensation for the smeared BCFT analogous to the Lieb-Shultz-Mattis theorem for noninvertible symmetry. Corresponding to the bulk topological degeneracies in $2+1$ dimensions and gapped quantum phases in $1+1$ dimensions, we construct a new series of BCFT. 

We also study the implications of the massless RG flow of $1+1$ dimensional CFTs corresponding to the domain wall or coupling problem in $2+1$ dimensional topological orders which corresponds to the earlier proposal by L. Kong and H. Zheng in [Nucl. Phys. B 966 (2021), 115384], arXiv:1912.01760 closely related to the integer-spin simple current by Schellekens and Gato-Rivera. We reveal that two distinct kinds of massless RG flow exist, coming from anomaly cancellation and anomaly matching related to Witt equivalence of bulk CFTs. The anomaly cancellation flows produce a new series of modular invariants as coupled models, and the anomaly matching flows produce nontrivial domain walls changing the chirality of anyons. Our paper gives a unified direction for the future theoretical and numerical studies of the topological phase based on the established data of classifications of conformal field theories or modular invariants (and the corresponding category theories).

\end{abstract}

\maketitle

\section{introduction}
\label{introduction}

Anyon condensation and Witt equivalence are the most fundamental structures classifying topological orders (TOs) or corresponding topological quantum field theories (TQFTs) in contemporary physics and related mathematics\cite{Bais:2008ni,Kong:2013aya,Burnell_2018}. The theories are deformed to each other without obstruction when the two theories are properly related by the classifications. Whereas the systematic studies on such classification schemes are still in development and sometimes contain controversies, the fundamental significance of abstract algebra of anyons and the corresponding category theories have already been established to some extent (see \cite{Simon:2023hdq,Kong:2022cpy,Kawahigashi:2021hds,Evans:2023nbp} and reference therein, for example). 
Combining these ideas with gauging (or group extension) and orbifolding which has been studied in the context of renormalization group (RG) flow\cite{Cappelli:1987xt,Cappelli:2009xj,Fuchs:1992nq}, one can expect some general correspondence between the classification of TOs in $2+1$ dimensions and that of the RG flows in $1+1$ dimensional CFTs\cite{Kaidi:2021gbs,Fukusumi_2022_c,Kikuchi:2024ibt,Fukusumi:2024nho,Kikuchi:2024cjd}. This is a modern version of bulk-edge correspondence or correspondence between conformal field theory (CFT) and TQFT, known as CFT/TQFT\cite{Witten:1988hf,moore_nonabelions_1991}. 

The symmetry algebraic or category-theoretic version of CFT/TQFT\cite{Fuchs:2002cm,Kong:2017etd,Kong:2019byq,Kong:2019cuu,Ji:2019jhk}\footnote{We thank Liang Kong  for notifying us a lot references and their status in the fields.}, called ``topological holography" \cite{Moradi:2022lqp}, has been studied widely recently (However, the terminology ``topological holography" appeared earier in a different contenxt in \cite{Husain:1998vz} representing systems satisyfying both topological and holographic property). The resultant sandwich construction\cite{kong2015boundarybulkrelationtopologicalorders,Kong:2017hcw,Bhardwaj:2023bbf} has been applied for several models\cite{Huang:2023pyk,Wen:2024udn,Bhardwaj:2024wlr,Fukusumi:2024cnl,Bhardwaj:2024ydc,Huang:2024ror,Vanhove:2024lmj,Lan:2024lis}. Based on these backgrounds, constructing a classification of CFTs is fundamentally important for the classification of TOs. However, we stress that in such generalized symmetry-based studies, the theoretical and mathematical framework on the famous TOs, fractional quantum Hall effects (FQHEs), have still been under development regardless of their experimental significance\cite{PhysRevLett.48.1559}. It might be surprising for readers, but the existing famous framework, modular-tensor-category (MTC) or bosonic chiral CFT (CCFT), does not correspond to the FQHEs in general. It is necessary to extend the bosonic theory by a finite group and this is called simple current extension\cite{Schellekens:1990ys,Gato-Rivera:1991bqv,Gato-Rivera:1991bcq,Kreuzer:1993tf,Fuchs:1996rq,Fuchs:1996dd}. When concentrating our attention on the $Z_{2}$ or fermionic case, such extended structure has already appeared in the established works \cite{moore_nonabelions_1991,Milovanovic:1996nj,Cappelli:1996np,Ino:1998by,Ino:2000wa,Ino:2000ve}. More generally, one can see the corresponding arguments widely in older theoretical literature \cite{Cappelli:1996np,Cappelli:1998ma,Frohlich:2000qs,Lu_2010,Cappelli:2010jv,Cappelli:2013iga,Schoutens:2015uia}. The resultant structure has captured attention of the fields as realizations of supersymmetry in condensed matter\cite{PhysRevLett.108.256807,Yang_2013,Sagi:2016slk,Gromov:2019cgu,Bae:2021lvk,Ma:2021dua,Fukusumi:2022xxe,PhysRevLett.130.176501,Nguyen:2022khy}. More recently, the nontriviality of the corresponding categorical frameworks has been studied in \cite{Lan_2016,Lan2016ModularEO,Ji:2019ugf}. The main difficulty in studying group extensions of CFTs is the treatment of (fermionic) zero modes \cite{Ginsparg:1988ui} (or semion\cite{Kitaev:2006lla}). This zero mode counting results in the degeneracies of wavefunctions of FQHEs\cite{moore_nonabelions_1991,Nayak_1996}, but the semion structure has not been studied extensively except for several earlier works\cite{Ginsparg:1988ui,Ino:1998by,Milovanovic:1996nj,Ardonne:2010hj,Aasen:2017ubm,Chen_2020} and the recent works by the author \cite{Fukusumi:2022xxe,Fukusumi_2022,Fukusumi_2022_c,Fukusumi:2023psx,Fukusumi:2024cnl}. This is in parallel to the recent development of the studies on fermionic or gauged models\cite{Gaiotto:2015zta,Bhardwaj:2016clt,Tachikawa:2017gyf,Bhardwaj:2017xup,Kobayashi:2019xxg,Inamura:2022lun,Debray:2023iwf,Zhou:2024fkv} (The references are huge, so we have only listed established literature or literature containing discussions on indicators).  
 The zero modes are closely related to the order-disorder duality of lattice models\cite{Seiberg:2023cdc}, and the group extension is a procedure to unify order and disorder operators canonically in QFTs. Hence, by including nonlocal objects, the theory after the extension becomes outside of MTCs. The resultant theory has been studied as a super-fusion category or premodular fusion category, for example, but this research direction is still in development.

\begin{figure}[htbp]
\begin{center}
\includegraphics[width=0.5\textwidth]{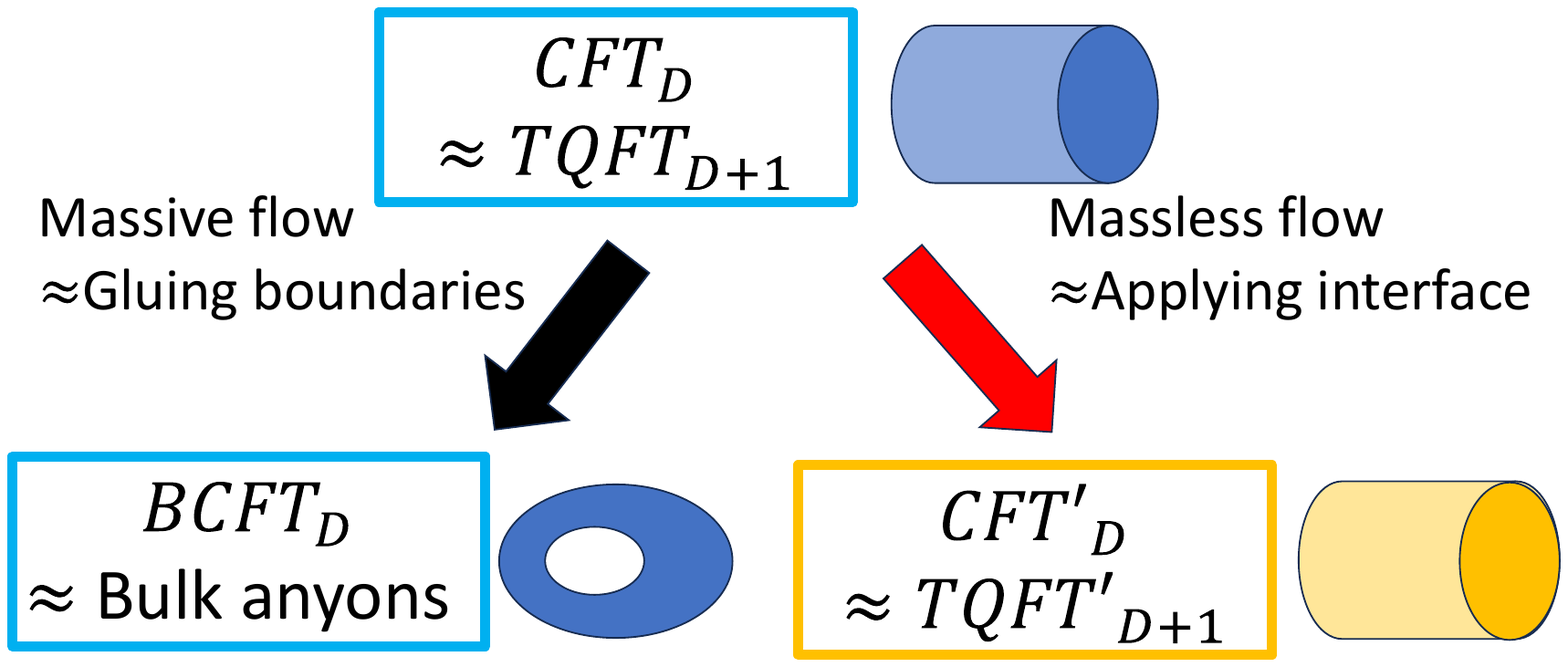}
\caption{Summary of the relationship between RG flow in CFTs and the notions in TQFTs. The upper figure shows $D$-dimensional TQFT with two boundaries. A FQH system in cylinder geometry is a typical example. The massive RG flow of CFTs or the (extended) smeared BCFTs\cite{Cardy:2017ufe} in this work determine the structure of bulk topological order with a close connection to the entanglement entropy\cite{Li_2008,Qi_2012,Lencses:2018paa,Das:2015oha,Lou:2019heg}. In TQFT, the  massive RG flow in CFTs corresponds to the gluing operation of the two boundaries. On the other hand, the massless RG flow of CFTs implies the (Witt) equivalence relation between group extended TQFTs (or extended SFCs\cite{Bischoff:2019jho,davydov2021braidedpicardgroupsgraded}). Applying the interface or domain wall known as a gapped or symmetry-preserving domain wall\cite{Kaidi:2021gbs} in TQFT corresponds to the massless RG flow. By analyzing $BCFT_{D}$ with $D\ge 3$ as an ancillary CFT \cite{Nishioka:2022ook}, a similar phenomenology will be true in general. We also note that the existing Witt equivalence for MTC (not SFC) seems too strong when applying to realistic settings\cite{Kawahigashi:2015lxa,Kawahigashi:2021hds,Moller:2024plb,Moller:2024xtt} (we stress that this does not mean their arguments are wrong. This implies the significance of extension for application to more realistic settings). One needs to introduce (or relax) the equivalence relation to the gauged bulk or chiral CFT. 
}
\label{TQFT_RG}
\end{center}
\end{figure}

When studying classifications of CFTs, there exist two fundamental aspects: symmetry and renormalization group (RG) flow. Recently, it has been gradually established that a RG flow to a gapped phase can be mapped to the classification of boundary conformal field theory (BCFT)\cite{Date:1987zz,Saleur:1988zx,Foda:2017vog,Cardy:2017ufe}, and a massless RG flow \cite{Zamolodchikov:1987ti,Zamolodchikov:1987jf,Zamolodchikov:1989hfa} can be mapped to the classification of the defect (or interface) of CFTs\cite{Crnkovic:1989ug,Brunner:2007ur,Gaiotto:2012np} (We also note the literature on the massless RG flows  \cite{Ravanini:1992fs,Ravanini:1992qb,Dorey:1992pj,Dorey:1992bq,Tateo:1994pb,Dorey:1996he,Dorey:2000zb,Sfetsos:2017sep,Georgiou:2017jfi}). In the context of CFT/TQFT, the former corresponds to the classification of bulk states or entanglement spectrum of TOs \cite{Li_2008,Qi_2012,Lencses:2018paa,Das:2015oha,Lou:2019heg} and the latter corresponds to the classification of boundary anyons induced by the Witt equivalence \cite{Kaidi:2021gbs,Fukusumi_2022_c,Kikuchi:2024ibt,Fukusumi:2024nho} (Fig. \ref{TQFT_RG}). Corresponding discussions for the TQFT side can be seen in the studies of the gapped domain wall in \cite{Lan:2013wia,Lan:2014uaa,Kawahigashi:2015lxa,Kong:2017etd,Kawahigashi:2021hds,Kong:2019byq,Huston:2022utd,Kong:2019cuu,Zhao:2023wtg} which can be interpreted as a particular class of symmetry-preserving domain walls in \cite{Kaidi:2021gbs,Fukusumi_2022_c}. 

\begin{figure}[htbp]
\begin{center}
\includegraphics[width=0.5\textwidth]{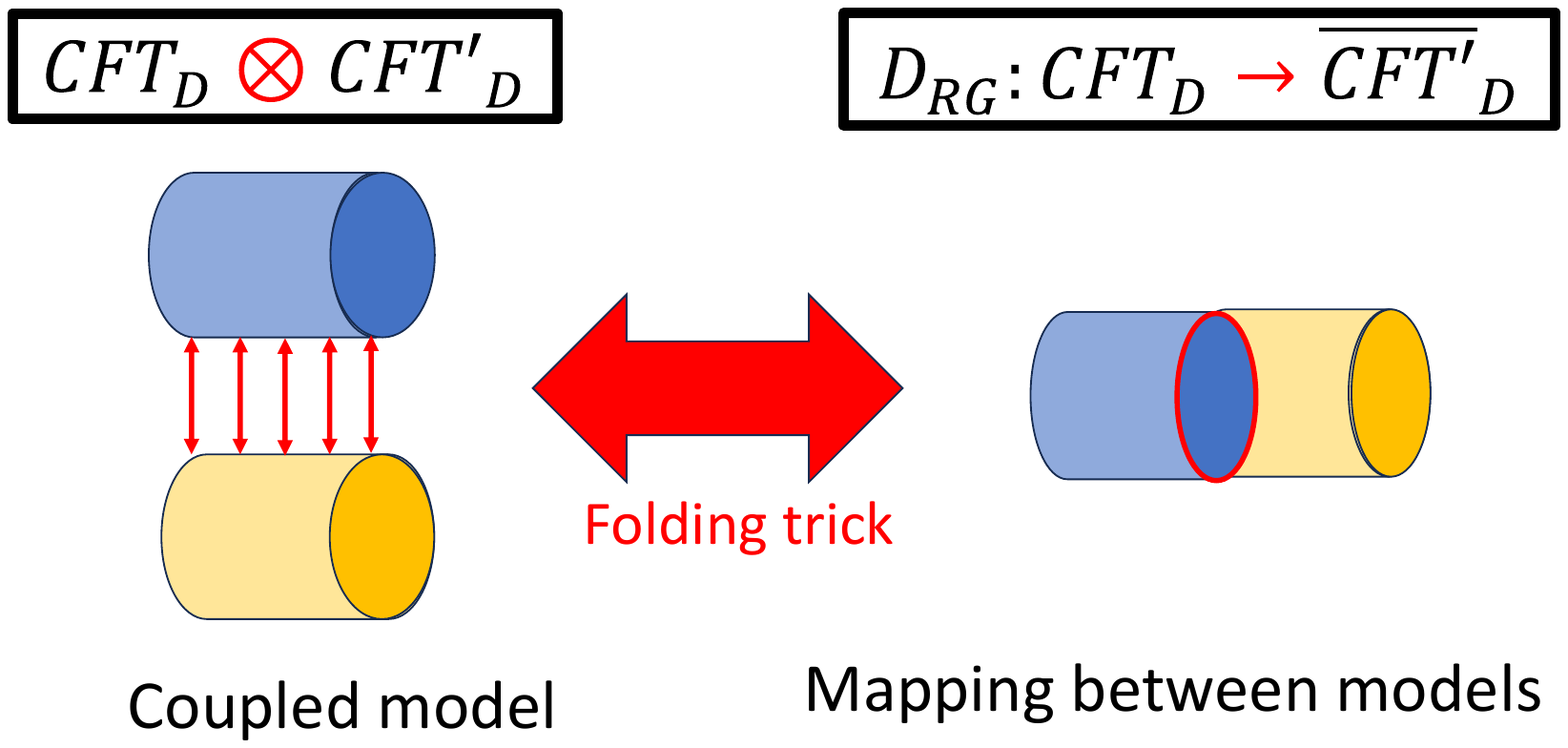}
\caption{Folding trick applied to coupled models. By applying the folding trick\cite{Wong:1994np}, one can obtain the correspondence between a coupled model ($CFT_{D}\otimes CFT'_{D}$) and anyon transformation law of two theories($D_{RG}: CFT_{D}\rightarrow \overline{CFT'_{D}}$). Similar arguments can be seen in \cite{Kaidi:2021gbs,Zhao:2023wtg}, for example. Because of the folding, the orientation or chirality of one of the two bulk CFTs is reversed. The coupled models in this work are the multicomponent or multilayered TOs. The mapping $D_{\text{RG}}$ corresponding to the symmetry-preserving domain wall\cite{Kaidi:2021gbs} or RG domain wall\cite{Brunner:2007ur,Gaiotto:2012np} classifies the TOs, possibly including FQHEs with nonchiral anyon. By shrinking one of the two systems, $CFT'_{D}$ in the righthand side of the figure, one can obtain the mismatch of bulk theory $CFT_{D}$ and boundary theory $CFT'_{D}$. This explains recent controversies on the thermal Hall conductance in experimental settings\cite{Mross_2018,Wang2017TopologicalOF}.}
\label{figure_folding_RG}
\end{center}
\end{figure}

The idea of RG domain wall is surprisingly concise because this is just attaching two systems connected under a massless RG flow \cite{Zamolodchikov:1987ti,Zamolodchikov:1987jf,Zamolodchikov:1989hfa} through a particular type of domain wall\cite{Crnkovic:1989ug,Gaiotto:2012np}. Hence, one can map the problem of RG to an algebraic problem of a domain wall (or conformal interface) between ultraviolet (UV) and infrared (IR) CFTs. Usually, the verification of RG requires extensive analytical calculations of quantum field theories (QFTs) and numerical tests in lattice models, and they tend to be difficult to formulate rigorously or mathematically. Moreover, by using a folding trick \cite{Wong:1994np} to the system divided by an RG domain wall, one can map the problem to the product of RG-connected CFT with a boundary\cite{Crnkovic:1989ug} and this is in the scope of BCFT analysis. Hence, one can apply techniques in defect or boundary CFTs rigorously and systematically in principle (Fig.\ref{figure_folding_RG}). Moreover, the validity of the RG domain wall in the minimal CFTs has been tested more rigorously in \cite{Cogburn:2023xzw}, and one can expect that the RG domain wall can provide a rigorous formulation of RGs. In this sense, further systematic studies on CFT with defects and boundaries are fundamentally important for the systematic classifications of TOs. We note recent works for the RG domain wall and related junction problems\cite{Quella:2006de,Brunner:2015vva,Kimura:2015nka,Stanishkov:2016pvi,Stanishkov:2016rgv,Klos:2019axh,Klos:2021gab,Poghosyan:2022mfw,Poghosyan:2022ecv,Zeev:2022cnv,Fukusumi:2023vjm,Konechny:2023xvo,Poghosyan:2023brb,Kikuchi:2022ipr,Kikuchi:2024cjd,Benedetti:2024utz}.

In this work, we systematically study algebraic properties of $2+1$ dimensional TOs as the RG flow of the corresponding two-dimensional CFT. Our formalism has several benefits compared with other existing frameworks. One can construct the theory or partition function only by using the data of existing $Z_{N}$ symmetric modular invariants (see \cite{Cappelli:1987xt,Cappelli:2009xj,Fuchs:1992nq} for example). They usually contain the information of central charge, conformal dimensions, and modular $S$ matrix, and one can see the type or $Z_{N}$ anomaly of anyons. For example, in the discussion for massless flow, our work provides general data of $Z_{N}$ anomaly classification of (chiral or nonchiral) anyons and this can be thought of as an extension of the pioneering work by Kong and Zheng \cite{Kong:2019cuu} and older work by Schellekens and Gato-Rivera\cite{Schellekens:1990ys,Gato-Rivera:1991bqv,Gato-Rivera:1991bcq,Kreuzer:1993tf,Fuchs:1996rq,Fuchs:1996dd} (We list the reviews by Fuchs and his collaborators for useful references\cite{Fuchs:1997af,Schweigert:2000ix}). In the older literature by Schellekens and Gato-Rivera, there seem to exist remarks corresponding to our method (or the situation considered by Kong and Zheng). However, the resultant partition functions are difficult to read. Moreover, as far as we have checked, the general case that we study has not been mentioned in the related literature\cite{Fuchs:1997af,Schweigert:2000ix}. Related arguments on nonchiral anyons in some particular TOs can be seen only in \cite{Fuji_2017,Sule:2013qla,Sohal_2020} \footnote{We also note the simplest cases for the $SU(2)$ models have been studied by Tokiro Numasawa, in his unpublished notes. We thank Tokiro Numasawa for sharing his notes with us.}. For the readers interested in the implications of this work in numerical and experimental settings in condensed matter, we remark on four phenomena or models, multilayer or multicomponent system\cite{Halperin:1983zz}, nonchiral anyon in the antiPfaffian and higher Landau level (HLL), and controversy of thermal Hall conductance\cite{Mross_2018,Wang2017TopologicalOF}, and Landau level mixing. They can be explained by using our formalism without inconsistencies at this stage. Respectively, quantum field theoretic explanations of such phenomena have already been proposed (see review \cite{Hansson_2017} and reference therein), but we believe our formalism is the most concise, based on the anomaly analysis and anyon physics (or a modern version of quark-hadron particle physics). Moreover, in the discussion of massive RG flow, we provide a new class of BCFT, $Z_{N}$ extended BCFT which determines the gapped phase of $1+1$ dimensional quantum lattice models and the bulk properties of the corresponding $2+1$ dimensional TOs. For example, we demonstrate a concise explanation of $1+1$ dimensional $Z_{2}$ symmetry-protected topological (SPT) phase. Hence, our study provides useful data on both bulk and boundary classifications of TOs.
 
Compared with the existing literature requiring knowledge of category theory, the requirement of our formalism is much simpler, the table of conformal dimension in a modular invariant. Hence, one can obtain the benefit of the established systematic construction of coset CFTs\cite{Goddard:1984hg,Goddard:1984vk,Goddard:1986ee}. We itemize the fundamental properties to check the data of CFTs or the corresponding modular forms,  
\begin{itemize}
\item{Anyons behaving like discrete group symmetry generators (or simple currents)}
\item{Symmetry charge (or anomaly) of anyons determined by the modular $S$ matrix}
\item{Conformal or scaling dimension of anyons}
\end{itemize}
The first point determines the fundamental structure of generalized electronic objects, such as Majorana, $Z_{N}  $ parafermion. The second point characterizes the stability of anyons, and the third point combined with the fusion algebra of anyons provides the asymptotic behavior of the wavefunctions\cite{Lu_2010,Schoutens:2015uia,Fukusumi:2023psx}. Combining methods in this work and the existing matrix-product-state (MPS) or tensor-network calculations \cite{Zaletel_2012,Estienne_2013,estienne2013fractionalquantumhallmatrix,Wu_2015,Vanhove:2018wlb,Chen:2022wvy,Ueda:2024jzj} and related earlier studies in mathematical physics\cite{Fuchs:2002cm}, a more general construction of TOs will become possible in principle.

The rest of the manuscript is organized as follows. In section \ref{CFT_symmetry}, we introduce the basics of the bosonic CFTs and their defects and symmetries from a modern perspective. We also provide the fusion algebra data of Ising and Majorana CFTs for later discussions. In section \ref{QFT/BCFT}, we discuss applications of symmetry-based arguments on the classification of massive RG flows from CFTs. We study the implications of noninvertible symmetries and the basic data of the new BCFT corresponding to both the spontaneous symmetry-breaking phase (SSB) and disordered phase. In section \ref{section_massless_flow_TQFT}, we introduce the improved expression of the anomaly-freeness in a CFT and the corresponding TQFT with emphasis on the correspondence between the domain wall and massless RG flow. A new series of modular invariant or domain walls involving nontrivial mixing of chiralities is shown. One can obtain sufficient data from a single modular partition function Eq. \eqref{universal_partition_function}. Section \ref{conclusion} is the concluding remarks. We summarize the implications of our formalisms and note general phenomenologies behind gauging procedures which will be useful in establishing a concrete understanding of related phenomena in physics. In the Appendix, we note several detailed systematic constructions of the models in the main texts with historical remarks.

\emph{Note for readers}: In this work, we have revisited studies in various research fields. To avoid confusion, we have restricted our attention to aspects of abstract algebra and modular form and related QFTs. The main concepts are summarized in the figures and itemized or labeled statements. We expect that they will be useful to readers unfamiliar with recent mathematical arguments.

\subsection{Summary of the main results}

In this manuscript, we provide the construction and analysis of BCFTs, coupled or multicomponent CFTs corresponding to a series of domain walls. Corresponding to these topics, we summarize the main (or new) results in this work.

In section \ref{CFT_symmetry} and \ref{QFT/BCFT}, we provide a general analysis of smeared BCFTs and their implications for $1+1$ dimensional gapped phase. A general condition forbidding the appearance of smeared Cardy states as gapped ground states is discussed. The most essential condition is summarized as Eq. \eqref{condensation_condition} relating the existence of noninvertible symmetry of CFT and condensation conditions in the resultant gapped phase. We also provide a new class of boundary states, eq.\eqref{smeared_SSB_1} and eq.\eqref{smeared_SSB_2}, which correspond to an SSB or smeared BCFT.

In section \ref{section_massless_flow_TQFT}, we provide a new series of multicomponent CFTs and corresponding domain walls. The most fundamental finding in this section is summarized in the form of a partition function, Eq. \eqref{universal_partition_function}. The phenomenologies are summarized in two figures, Fig. \ref{CFT_OECFT} and Fig. \ref{anyon_transformation}, respectively. For the readers interested in the realization of the multicomponent model, we demonstrate a construction in $1+1$ dimensional quantum lattice models in Appendix \ref{Lattice_realization} and interpretations of the corresponding wavefunctions in Appendix \ref{nonchiral_wavefunction}. One can obtain a more intuitive understanding of the nontriviality of the mixing of chirality from Fig. \ref{Dirac_type_parefermion}.

\section{Boundary and defect conformal field theories: chiral extension, symmetry charge and amplitude}
\label{CFT_symmetry}
In this section, we introduce a short review of CFTs with boundaries and defects and the notions relevant to the successive discussions. We note reviews \cite{Cardy:2004hm,Petkova:2000dv} and a textbook  \cite{Recknagel:2013uja} as references for boundary CFTs and \cite{Petkova:2000ip} for defect CFTs. We also note recent lecture notes containing both of the topics \cite{Northe:2024tnm}, and the standard textbook of CFTs \cite{DiFrancesco:1997nk}. For the readers interested in the recent problems in the fields, we note \cite{Andrei:2018die}.

The starting point of the discussion in this manuscript is the bosonic bulk CFT with a diagonal partition function,
\begin{equation}
Z=\sum_{\alpha}\chi_{\alpha}(\tau)\overline{\chi}_{\overline{\alpha}} (\overline{\tau})
\end{equation}
where $\alpha$ is the label of (bosonic) chiral primary fields and $\chi$ and $\overline{\chi}$ are the chiral and antichiral characters respectively and $\tau$ and $\overline{\tau}$ are the corresponding modular parameters. In this work, the Greek symbols, $\alpha, \beta, ...$ are labels of fields without respective remarks.

To discuss a boundary and defect CFT, we introduce the following modular $S$ transformation,
\begin{equation} 
\tau \rightarrow \frac{-1}{\tau}, 
\end{equation}
The matrix representation of the modular $S$ transformation is,
\begin{equation} 
\chi_{\alpha}\left(\frac{-1}{\tau} \right)=\sum_{\alpha'} S^{\alpha'}_{\alpha} \chi_{\alpha'} (\tau).
\end{equation}
The modular $S$ transformation governs the open-closed (or low and high temperature) duality\cite{Cardy:1989ir}. This is a phenomenological reason why the entanglement spectrum of a single ground state of TO knows its total edge spectrum\cite{Li_2008,Qi_2012,Lencses:2018paa,Das:2015oha,Lou:2019heg}. 
For the later discussion, we introduce the modular $T$ transformation as,
\begin{equation} 
\tau \rightarrow \tau +2\pi 
\end{equation}
This governs the locality or statistics of the theories.

For simplicity, we restrict our attention to a unitary model with $S^{\alpha}_{0}>0$ for all $\alpha$ where $0$ represents the vacuum of CFT. (However, one can apply a similar analysis for a class of nonunitary minimal models by replacing the exact vacuum with the effective vacuum. This exchange corresponds to the Galois shuffle in \cite{Gannon:2003de,Harvey:2019qzs,Fukusumi_2022,Fukusumi_2022_c}.)

The Cardy states, which correspond to a class of boundary conditions describing the boundary phenomena in $1+1$ dimensional lattice models, can be represented as,
\begin{equation}
|\alpha\rangle \rangle =\sum _{\alpha'}\frac{S^{\alpha}_{\alpha'}}{\sqrt{S^{\alpha}_{0}}} |\alpha\rangle\rangle_{\text{I}}
\end{equation}
where the righthand side corresponds to Ishibashi states\cite{Ishibashi:1988kg}. To avoid complicated notations unnecessary for the successive discussions, we only note the basic property of Ishibashi states,
\begin{equation}
\langle\langle \alpha|_{\text{I}} e^{i\pi \tau H_{\text{CFT}}}|\alpha'\rangle\rangle_{\text{I}}=\delta_{\alpha,\alpha'}\chi_{\alpha}(\tau)
\end{equation}
where the $H_{\text{CFT}}$ is the Hamiltonian of the full CFT and $\delta$ is the Kronekker delta. The Cardy states combined with the Verlinde formula\cite{Cardy:1989ir,Verlinde:1988sn} provide the following amplitude,
\begin{equation}
\langle \langle \alpha| e^{i\pi \tau H_{\text{CFT}}}|\alpha'\rangle\rangle=\sum_{\gamma} n^{\gamma}_{\alpha \alpha'}\chi_{\gamma}(-1/\tau)
\end{equation}
where $n$ is the fusion matrix. For the readers in condensed matter, we note that the fusion matrix is the most fundamental structure to determine the algebraic property of anyons and corresponding generalized symmetries. We also note that one can recover the Ishibashi states from the Cardy states by using the linear relation between them. In other words, whereas this might be a bit unusual compared with the literature, it is phenomenologically more natural to start the discussion of BCFT from the fusion algebra and Cardy states.

From the arguments in \cite{Petkova:2000ip} the topological symmetry operator (or Verlinde line), which can be interpreted as a proto type of generalized global symmetry \cite{Cobanera:2009as,Gaiotto:2014kfa}, can be written as,
\begin{equation}
\mathcal{Q}_{\alpha}= \sum_{\alpha',m,\overline{m}} \frac{S^{\alpha'}_{\alpha}}{S^{0}_{\alpha}} P_{\alpha',\overline{\alpha'}, m,\overline{m}}
\label{Verlinde_line}
\end{equation}
where $m$ and $\overline{m}$ is the label of descendant states and $P$ is the corresponding projection operator. Because these operators are written as a summation of projections, they commute with the Hamiltonian and behave as symmetries (or conserved quantities). The topological symmetry operator can become noninvertible and this has captured attention recently in the classification of TOs\cite{Ji:2019jhk}. If one interprets $\mathcal{Q}$ as a symmetry (or integral of motion), the phase and amplitude of $S^{\alpha'}_{\alpha}/S^{0}_{\alpha}$ play a fundamental role. For this purpose, we introduce the following quantity,
\begin{equation}
\frac{S^{\alpha'}_{\alpha}}{S^{0}_{\alpha}}=r_{\alpha \alpha'}e^{i2\pi i Q_{\alpha}^{\text{mon}}(\alpha')}
\end{equation}
where $r$ is a real number with $r\ge 0$ and $Q_{\alpha}(\alpha')$ defined by modulo $1$ is the charge of the quantum states labeled by $\alpha'$ under the symmetry $\mathcal{Q}_{\alpha}^{\text{mon}}$.  
For example, when $\mathcal{Q}_{\alpha}$ corresponds to a $Z_{N}$ symmetry, the amplitude produces $r_{\alpha \alpha'}=1$ and the phase produces the $Z_{N}$ symmetry charge or anomaly. Hence it is reasonable to name  $r_{\alpha \alpha'}$ as symmetry amplitude for the quantum state labeled by $\alpha'$ under symmetry $\mathcal{Q}_{\alpha}$. Moreover, when the symmetry is noninvertible, some matrix elements of $S$ and the corresponding amplitudes should vanish. In Section \ref{QFT/BCFT}, we study the implications of the noninvertible symmetry from this view. 

\subsection{Algebraic data of Ising and Majorana CFT: Bulk and chiral}

In this subsection, we introduce the fusion algebraic data of Ising or Majorana CFT for later use. We cite \cite{Ginsparg:1988ui,DiFrancesco:1997nk} as basic references and \cite{Ji:2019jhk,Huang:2023pyk,Inamura:2023ldn,Wen:2024udn,Fukusumi:2024cnl,Huang:2024ror,Bhardwaj:2024ydc} for the recent works containing related arguments. First, the Ising CCFT has three objects (or chiral primary fields) $\{ I,\psi,\sigma\}$ with the respective conformal dimension $\{ 0, 1/2,1/16 \}$ satisfying the following algebraic relations known as MTC,
\begin{align}
\psi\times \psi&=I, \\
\psi\times \sigma&=\sigma, \\
\sigma \times \sigma&=I+\sigma. 
\end{align} 
The object $\psi$ is called chiral Majorana fermion. Because of its property as a $Z_{2}$ symmetry generator, $\psi$ is also called $Z_{2}$ simple current. The relevant point for the later discussions is the $Z_{2}$ invariance of the object $\sigma$. Because of this structure, one cannot map the above algebra to $Z_{2}$ group ring nontrivially. Technically, this can be understood as an obstruction for $Z_{2}$ grading or fermion parity attachment (or introduction of $Z_{2}$ particle number counting).

By the ring isomorphism or anyon condensation\cite{Bais:2008ni,Kawahigashi:2021hds} equivalent to the Moore-Seiberg data\cite{Moore:1988qv,Moore:1988ss}, one can obtain the algebraic representation of bulk primary fields known as a spherical fusion category (SFC) or nonchiral fusion rule \cite{Rida:1999xu,Rida:1999ru,Balaska:2002qw,Balaska:2003xx}(In this work, we use the term ``nonchiral" in a different way, hence we use SFC except for here), 
\begin{align}
\epsilon\times \epsilon&=I, \\
\epsilon\times \sigma_{\text{Bulk}}&=\sigma_{\text{Bulk}}, \\
\sigma_{\text{Bulk}} \times \sigma_{\text{Bulk}}&=I+\epsilon, 
\end{align} 
where the identification from the ring isomorphism is $\{ I,\psi, \sigma\} \sim \{ I, \epsilon, \sigma_{\text{Bulk}}\}$. $I$ is the identity operator, $\epsilon$ is the energy and $\sigma_{\text{Bulk}}$ is the $Z_{2}$ order operator operator.
For later discussions, we stress that we do not assume the chiral and antichiral decomposition without respective cautions, because this decomposition sometimes contains subtelities. Moreover, in general, this is not a tensor decomposition (in category theory, this is called Deligne product \cite{Deligne1990} and this structure is verified by ``abstract nonsense"\cite{Kong:2009inh,Kong:2013aya}). In any case, the above three algebraic objects reproduce the established bosonic modular invariant $|\chi_{I}|^{2}+|\chi_{\psi}|^{2}+|\chi_{\sigma}|^{2}$.

By the Jordan-Wigner transformation or $Z_{2}$ simple current extension\cite{Hsieh:2020uwb}, the bulk primary fields of Majorana CFT have six objects $\{I,\psi, \overline{\psi}, \epsilon, \sigma_{\text{Bulk}}, \mu_{\text{Bulk}} \}$ with the following algebraic relations,
\begin{align}
\psi\times \psi=\overline{\psi} \times \overline{\psi}&=\epsilon \times \epsilon=I, \\
\epsilon\times \sigma_{\text{Bulk}}&=\sigma_{\text{Bulk}}, \\
\psi\times \sigma_{\text{Bulk}}=\overline{\psi}\times \sigma_{\text{Bulk}}&=\mu_{\text{Bulk}}, \\
\sigma_{\text{Bulk}} \times \sigma_{\text{Bulk}}&=I+\epsilon, \\
\mu_{\text{Bulk}} \times \mu_{\text{Bulk}}&=I+\epsilon,
\end{align} 
Because this is an extension (or gauging), the number of bulk fields becomes twice. Roughly speaking this extension has been achieved by the introduction of the chiral (antichiral) $Z_{2}$ field $\psi$ ($\overline{\psi}$) and its multiplication to the Ising SFC. The most nontrivial point of this extension is the introduction of the disorder fields $\mu_{\text{Bulk}}$ represented as $\mu_{\text{Bulk}}=\psi \times \sigma_{\text{Bulk}}$, because the naive tensor product representation $\sigma \otimes \overline{\sigma}$ results in the relation $\psi \times (\sigma \otimes  \overline{\sigma})=\sigma \otimes \overline{\sigma}$. Because of the extension, only the Neveu-Schwarz sector $\{ I,\psi, \overline{\psi}, \epsilon \}$ forms the modular $S$ invariant $|\chi_{I}+\chi_{\psi}|^{2}$ and the locality is reduced to modular $T^{2}$ invariance. The other sector, $\{ \sigma_{\text{Bulk}}, \mu_{\text{Bulk}}\}$ known as the Ramond sector only forms modular covariant $2|\chi_{\sigma}|^{2}$.

By applying the anyon condensation or sandwich construction to this theory\cite{Ji:2019jhk,Huang:2023pyk,Wen:2024udn,Fukusumi:2024cnl,Huang:2024ror,Bhardwaj:2024ydc}, one can obtain the chiral Majorana CFT with the four objects $\{ I,\psi, e, m\}$ with respective conformal dimensions $\{ 0,1/2, 1/16, 1/16 \}$ (The algebraic details have been summarized in the author's work\cite{Fukusumi:2024cnl}). The algebraic data of this CCFT corresponds to those of the SymTFT (or categorical symmetry of $2+1$ dimensional TOs). For the readers interested in the classification of SymTFTs, we note recent works \cite{Bhardwaj:2024qiv,Bullimore:2024khm,Bhardwaj:2025piv,Bhardwaj:2025jtf}.

The algebraic relation of this SymTFT is called double-semion algebra, 
\begin{align}
\psi\times \psi=e\times e&=m\times m=I, \\
\psi\times e&=m, \\
\psi \times m&=e. 
\end{align} 
Because of the $Z_{2}$ extension, the $Z_{2}$ invariant object $\sigma$ split to the semions $\{ e, m\}$. Phenomenologically, these semions correspond to the zero modes of the Ramond sectors and result in structures with nontrivial entanglement\cite{Stern2003GeometricPA,Fukusumi:2023psx}. Whereas the semions behave as the $Z_{2}$ symmetry generators, their conformal dimensions are $1/16$. Hence, the semions are anomalous in the $Z_{2}$ classification in the sense of spin-statistics\cite{Pauli:1940zz} and this leads to nontrivial entanglement or nonlocality in the theory\cite{Stern2003GeometricPA,Fukusumi:2023psx}. As we will discuss in the next section, this nontrivial entanglement structure has a close connection to the classification of $Z_{2}$ SPT\cite{Furuya:2015coa,MOshikawa_1992,Pollmann:2009ryx,Pollmann:2009mhk}.

After the $Z_{2}$ extension, one can define $Z_{2}$ grading for each particle in the CCFT. For example, one can take $I$ and $e$ as $0$, and $\psi$ and $m$ as $1$. One can check its consistency by replacing each object in the double semion algebra to $0$ or $1$ and $\times$ to $+$,
\begin{align}
1+1=0+0&=1+1=0, \\
1+0&=1, \\
1 +1&=0. 
\end{align} 
This nontrivial replacement is impossible for the original Ising MTC. In other words, the Majorana CCFT is the better basis to apply for the phenomenology of condensations which have a root in particle physics (or quark-hadron physics\cite{Goddard:1984vk,Goddard:1984hg}). We also point out the close relationship between this kind of phenomenology and the BCS theory\cite{Bardeen:1957mv}.

\section{Ordering by bulk relevant perturbation and its interpretation in BCFT}
\label{QFT/BCFT}

In this section, we revisit the conjecture by Cardy \cite{Cardy:2017ufe} with emphasis on a modern understanding of (generalized) symmetry. For simplicity, we concentrate our attention on the system corresponding to a $1+1$ dimensional quantum many-body system or lattice model with a periodic boundary condition. By applying the open-close duality to the resultant smeared BCFT, one can obtain the CCFT and corresponding TQFT. The phenomenology can be summarized as,
\begin{equation}
\{ \text{CFT/TQFT} \}\overset{\text{RG}/\text{Gluing}}{\Rightarrow} \{\text{BCFT/CCFT}\}
\end{equation}

Let us start from the argument of the lattice model described by the following Hamiltonian $H_{\text{Lat}}\sim H_{\text{QFT}}$,
\begin{equation}  
H_{\text{QFT}}=H_{\text{CFT}}+H_{\text{p}}
\end{equation}
where $H_{\text{CFT}}$ is the CFT Hamiltonian and $H_{\text{p}}$ is the perturbations described by fields in the CFT. By choosing $H_{\text{p}}$ accordingly, one can describe the respective quantum phase transition (Numerically, this CFT-based description has been studied by using the truncated conformal space approach (TCSA)\cite{Yurov:1989yu,Yurov:1991my} often combined with the thermodynamic Bethe ansatz). The massive RG flows of the above Hamiltonian correspond to the bulk states of TQFT by assuming the cut and gluing procedure of TQFT\cite{Li_2008,Qi_2012,Lencses:2018paa,Das:2015oha,Lou:2019heg}.

The conjecture by Cardy in \cite{Cardy:2017ufe} is simple and general (The correspondence between massive RG and BCFT has first appeared in the study of integrable models in our knowledge\cite{Date:1987zz,Saleur:1988zx,Foda:2017vog}). The conjectue says that the ground state of $H_{\text{Lat}}$, $|GS\rangle,$ can be approximated by smeared Cardy states $\sum_{\alpha}e^{-\tau_{\alpha}H_{\text{CFT}}}|\alpha\rangle$ constructed from Cardy states $\{ |\alpha\rangle\}$. The parameter $\tau_{\alpha}$ is a model parameter that depends on the finite size scaling analysis. However, as we will demonstrate, the contribution from $\tau_{\alpha}$ is not relevant to our purpose and we dropped the contribution for simplicity. For the readers interested in more detailed calculations, we note several recent works\cite{Lencses:2018paa,Kikuchi:2022ipr,Cordova:2022lms,Konechny:2023xvo}, 

This may be unfamiliar to readers, but one can understand this conjecture by considering the phenomenology of ordering.
\begin{itemize}
\item{Ordering induces fixing of some fields. In an ordered phase, some order fields are fixed. In the disordered states, disorder fields are fixed.}
\item{Boundary condition in BCFT induces fixing of fields as boundary effects. In Dirichlet boundary conditions, the order fields are fixed. In the Neumann boundary condition, the disorder fields are fixed.}
\end{itemize}
The nonlocal observable corresponding to the former is studied in the name, indicator, or topological invariant\cite{1985AnPhy.160..343K}. The point is that the observable that vanishes at the exact critical point or UV takes a finite value at IR. More specifically, the ordering in $H_{\text{QFT}}$ can be summarized as,
\begin{equation}
\{\langle0|\Phi_{\alpha} |0\rangle=0 \}_{\text{UV}} \overset{\text{RG}}{\Rightarrow} \{\langle GS|U_{\alpha} |GS\rangle \neq 0 \}_{\text{IR}}
\label{RG_condensation}
\end{equation}
where $\alpha$ is a label of bulk fields for the respective RG flow, $|0\rangle$ is the vacuum state in the CFT, $\Phi$ corresponds to bulk primary fields in the original CFT and $U$ is the corresponding Lieb-Shultz-Mattis (LSM) operator \cite{Lieb:1961fr,Schultz:1964fv,MOshikawa_1992,Cho:2017fgz,Yao:2018kel,Wamer:2019oge,Lecheminant:2024apm}. The operator $U$ generates the lattice analog of the elementary particle excitation in the original CFT. We have implicitly assumed the large system size limit, but one can apply the finite scaling analysis for this kind of argument numerically. The ground state expectation value of the LSM twist operator\cite{Resta:1994zz,Aligia_1999,Aligia_2005,Kobayashi:2018mnk,Furuya_2019,M_rquez_2024} is a corresponding example. More specifically, the value $\text{log} \langle GS|U_{\alpha} |GS\rangle$ can be considered as a topological invariant.

\begin{figure}[htbp]
\begin{center}
\includegraphics[width=0.5\textwidth]{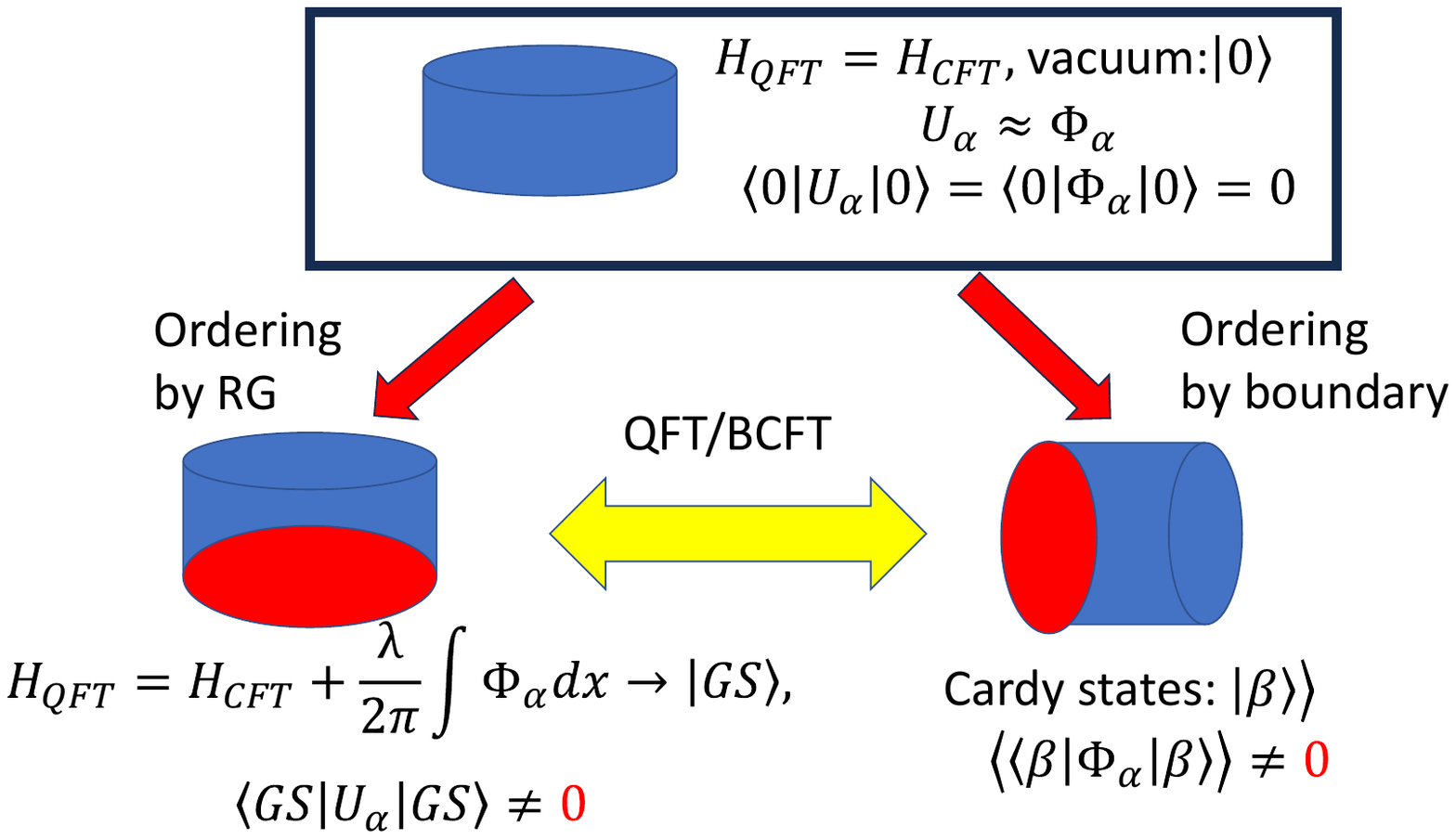}
\caption{ Mass-condensation or ordering in BCFT and RG. In a CFT without boundary, the vacuum expectation value of a bulk primary field $\Phi_{\alpha}$ should vanish. Under the operator-state correspondence, this vanishing corresponds to the vanishing of the expectation value of the LSM twist operator $U_{\alpha}$ corresponding to the many-body elementary excitation. Under RG, the ground state expectation value of the LSM operator becomes finite and this is a mass-condensation in the QFT side. The same condensation can be mimicked by introducing boundary to the same CFT, and this represents (massive) QFT/BCFT correspondence summarized by Cardy\cite{Cardy:2017ufe}.}
\label{mass_condensation_BCFT}
\end{center}
\end{figure}

The boundary condition induces the following condensations\cite{Cardy:2017ufe},
\begin{equation}
\langle\langle \beta|\Phi_{\alpha} |\beta\rangle\rangle \sim \frac{S^{\beta}_{\alpha}}{S^{0}_{\beta}} \left\{ \frac{S^{0}_{0}}{S^{0}_{\alpha}}\right\}^{1/2}
\end{equation}
where $|\beta\rangle \rangle$ corresponds to Cardy's boundary condition and we have omitted complicated scaling factors for simplicity (We also assumed the theory is diagonal).

One can represent the condensation induced by the effect of boundary as follows,
\begin{equation}
\{\langle0|\Phi_{\alpha} |0\rangle=0 \}_{\text{UV}} \overset{\text{B.C.}\beta}{\Rightarrow} \left\{\langle\langle \beta|\Phi_{\alpha} |\beta\rangle\rangle \sim \frac{S^{\beta}_{\alpha}}{S^{0}_{\beta}} \left\{ \frac{S^{0}_{0}}{S^{0}_{\alpha}}\right\}^{1/2}\right\}
\label{boundary_condensation}
\end{equation}

For convenience, we note the following representation,
\begin{equation} 
\frac{S^{\beta}_{\alpha}}{S^{0}_{\beta}} \left\{ \frac{S^{0}_{0}}{S^{0}_{\alpha}}\right\}^{1/2}=r_{\beta \alpha}e^{i2\pi i Q_{\beta}(\alpha)}d_{\alpha}^{-1/2}
\end{equation}
where $d_{\alpha}=S^{0}_{\alpha}/S^{0}_{0}$ is the quantum dimension of the object $\alpha$. The most fundamental point of this representation is that the condensation equivalent to the nonvanishing of operators is governed by the element of modular $S$ matrix, $S^{\beta}_{\alpha}$ or its amplitude $r_{\beta \alpha}$.

Comparing Eq. \eqref{RG_condensation} and Eq. \eqref{boundary_condensation}, one can map the problem of bulk RG flow to the classification of boundary conditions in BCFT and this is a concise representation of the proposal by Cardy \cite{Cardy:2017ufe} (Fig. \ref{mass_condensation_BCFT}). In other words, the vanishing of the element of the modular $S$ matrix restricts RG flows or the type of gapped phase determined by the mass condensation. The appearance of zero in the modular $S$ matrix elements implies obstruction for the condensation. Moreover, we note that the appearance of zero in the elements of a modular $S$ matrix results in the appearance of a noninvertible symmetry because of the matrix representation of the Verlinde line or Eq. \eqref{Verlinde_line} as a topological symmetry operator. Hence, one can obtain the following correspondence,
\begin{equation}
\text{Noninvertible symmetry} \sim \text{Obstruction to condensation}.
\end{equation}

By mimicking the argument of the LSM theorem, one can state ``if a resultant gapped ground state is described by a smeared Cardy state $|\beta\rangle \rangle$ with $S^{\alpha}_{\beta}=0$, the (nonzero) quantum state $\Phi_{\alpha}|\beta\rangle \rangle$ becomes other groundstates or gapped quantum states". The case where the quantum state $\Phi_{\alpha}|\beta\rangle \rangle$ corresponds to the ground state may be interesting because it implies a nontrivial relation between another Cardy state and the state $\Phi_{\alpha}|\beta\rangle \rangle$.

We study a more specific situation. Let us assume the form of the perturbation as $ H_{p}=(\lambda/2 \pi )\int dx (\Phi_{\alpha}+\Phi_{\alpha}^{\dagger})$ or $ H_{p}=(i\lambda/2 \pi )\int dx (\Phi_{\alpha}-\Phi_{\alpha}^{\dagger})$ with $S^{\alpha}_{\beta}=0$, where $x$ is the space cordinate and $\lambda$ is a coupling constant and $\dagger$ represents the Hermitian conjugate. In this setting, if the perturbation induces the fixing $\langle GS| \Phi|GS\rangle\neq 0$, this prevents the flow $|GS\rangle \rightarrow |\beta \rangle \rangle$  by contradiction. By identifying the smeared Cardy state $|\beta\rangle$ as a ground state of a gapped phase, we can obtain the following statement;

\emph{If  $r_{\alpha\beta}=0$ the system cannot flow to the gapped phase $|\beta\rangle\rangle$ under the perturbation $\Phi_{\alpha}$(and its Hermitian conjugate). More generally, the other perturbations inducing $\langle GS| \Phi_{\alpha}|GS\rangle\neq 0$ cannot flow the system to the gapped phase $|\beta\rangle\rangle$.} For the readers intersted in nontrivial relations between noninvertible symmetry and RG flows, we note another recent work by the author\cite{Fukusumi:2025clr}.

Formally, one can represent it as
\begin{equation}
\{|0\rangle, \text{with} \ r_{\alpha\beta}=0 \}_{\text{UV}} \overset{\text{RG},\langle\Phi_{\alpha}\rangle\neq 0}{\Rightarrow} \{ |GS\rangle \neq |\beta \rangle \rangle \}_{\text{IR}}.
\label{condensation_condition}
\end{equation}
This implies the following phenomenology,
\begin{equation}
\text{Noninvertible symmetry} \sim \text{Obstruction to RG flow}.
\end{equation}

Hence, one can state that the zero of the symmetry amplitude can restrict the phase diagram only by assuming RG flows from CFTs. This might be thought of as a noninvertible analog of symmetry-protected RG flow, but we have concentrated our attention on the symmetry amplitude, not a charge. If one induces the noninvertible symmetry $\mathcal{Q}_{\alpha}$ with $r_{\alpha\beta}=0$ as the integral of motion in the usual sense, this prohibits the existence of the perturbation $\Phi_{\beta}$. However, it should be mentioned that the vanishing of the amplitude $r_{\alpha\beta}=0$ does not necessarily mean $\mathcal{Q}_{\alpha} \Phi_{\beta} =0$ in considering lattice analog, because $\mathcal{Q}_{\alpha} \Phi_{\beta}$ can be a disorder operator by considering the commutation relation between $\mathcal{Q}_{\alpha}$ and $\Phi_{\beta}$(See \cite{Seiberg:2023cdc,Seiberg:2024gek}). Because a disorder operator is out of the scope of the local operators in the original bosonic CFT, the amplitude becomes zero. Inversely, the vanishing of the amplitude may imply the existence of a disorder operator, but this may require more careful arguments on the object $\mathcal{Q}_{\alpha} \Phi_{\beta}$. Hence, the following conjectural phenomenology will be important for future studies, 
\begin{equation}
\text{Existence of a disorder field} \ \Rightarrow \text{Obstruction to RG flow}.
\end{equation}
Similar discussions can be seen in \cite{Verresen:2019igf}. The precise relationship between the above arguments and those of defect RG flows in \cite{Chang:2018iay,Graham:2003nc,Kormos:2009sk} is also a future problem.

\begin{figure}[htbp]
\begin{center}
\includegraphics[width=0.5\textwidth]{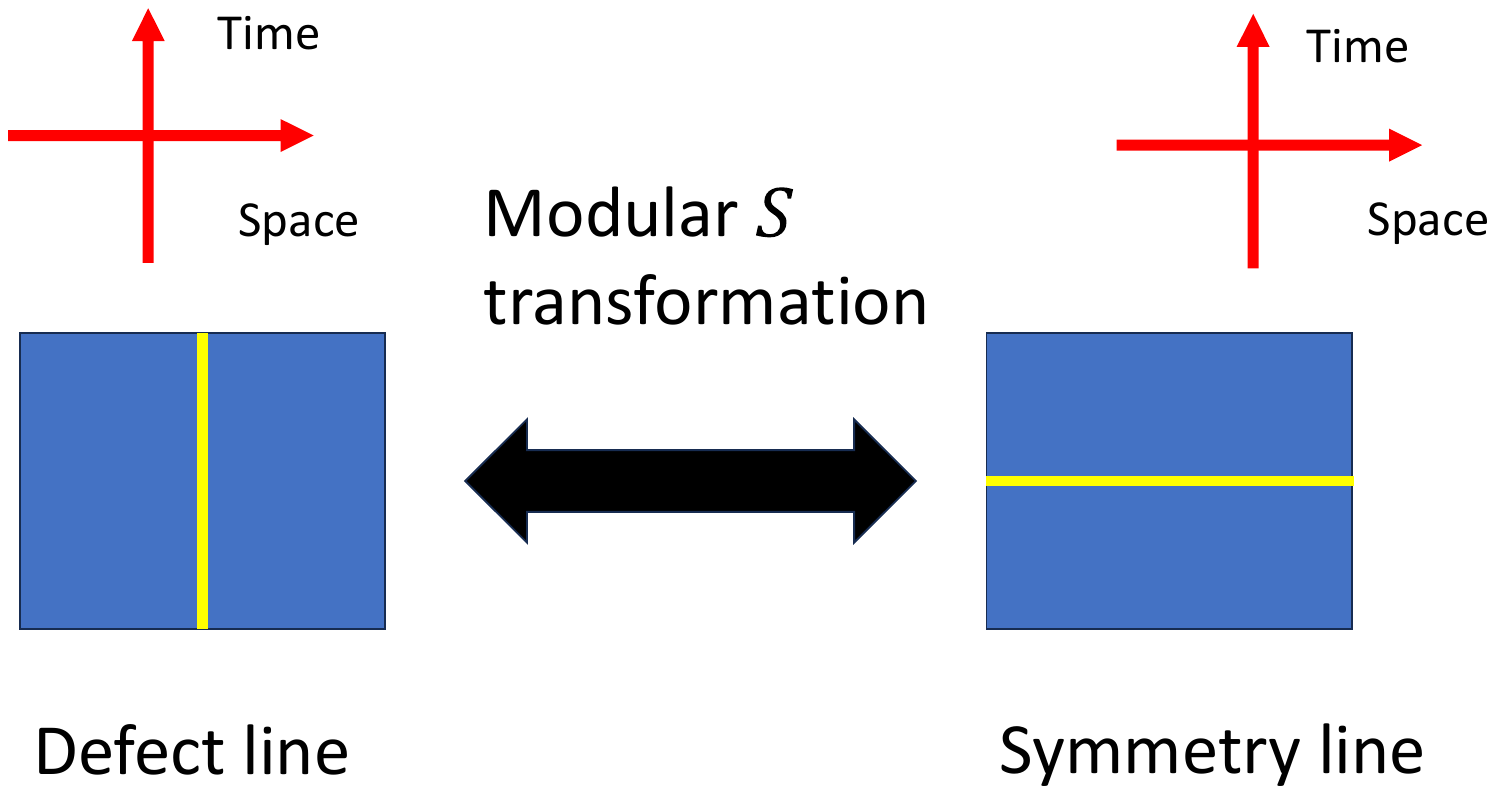}
\caption{The connection between symmetry and defect under modular $S$ transformation. The bondary states $|\beta\rangle\rangle$ can also be considered as defect condensation $\mathcal{Q}_{\beta}|0\rangle\rangle$\cite{Graham:2003nc,Fukusumi:2020irh}. Hence, in a large class of models, the combination of defects and symmetry governs the RG to the gapped phase.}
\label{space_time_defect}
\end{center}
\end{figure}

Considering the Lorentz invariance of the theory, it is useful to study the relationship between $\{\mathcal{Q}_{\alpha} \Phi\}$ and twisted theory by $\{ \mathcal{D}_{\alpha}\Phi\}$ where $\mathcal{D}_{\alpha}$ is a topological defect or domain wall (Fig.\ref{space_time_defect}). The action of $\mathcal{D}_{\alpha}$ to the bulk fields $\Phi$ can be implemented by the chiral and antichiral decomposition of $\Phi$, roughly speaking. Hence, by applying the modular $S$ transformation to the bulk fields, one can obtain the correspondence as follows. 
\begin{equation}
S: \{ \mathcal{Q}_{\alpha}\Phi \} \leftrightarrow \{ \mathcal{D}_{\alpha}\Phi \}
\end{equation}
We denote more detailed calculations in Appendix \ref{defect_symmetry_algebra}. When restricing the form of $\mathcal{D}_{\alpha}$ as $Z_{N}$ symmetry $J$, the exteded theory $\{\oplus_{i=0}^{N-1} \mathcal{D}_{J^{i}}\Phi_{\beta}\}$ corresponds to the simple current extension in the literature.

We note here the application of analysis in this section to the Ising CFT and discuss its interpretation in the quantum spin chain. There exist three chiral primary fields $\{ I,\psi, \sigma\}$, and one can label the corresponding Cardy states by them. $I$ and $\psi$ exchange each other, and they correspond to the ordered phase breaking $Z_{2}$ symmetry. The perturbation that triggers this flow is the primary fields $\sigma_{\text{Bulk}}$ and the $Z_{2}$ spin flip changing the sign of the coupling constant exchange each phase. The boundary state $|\sigma\rangle\rangle$ has to correspond to the invariant phase under the $Z_{2}$ spin flip. Hence this corresponds to the disordered phase. We remark that the $Z_{2}$ invariance of $\sigma$ plays a significant role in this identification. 

In the above arguments, the SSB phase is outside of the classification straightforwardly. We address more general cases including recent parafermion or fractional supersymmetric models. We also note the existing formalism using (weak) Hopf algebra for the massive integrable field theory known as Yangian symmetry\cite{Bernard:1990jw,Bernard:1992ya}, hidden quantum group symmetry\cite{Alvarez-Gaume:1988bek,Moore:1989ni,Gomez:1989sw,Felder:1991hw,Mack:1991tg,Gomez:1992bj} and the similar analysis for the SPT phases\cite{Inamura_2021,Inamura:2021szw,Jia:2024zdp,Meng:2024nxx}. The relationship between chiral algebra, BCFTs and quantum group (and the resultant Hopf algebra structure) have been appeared several times in different fields by using different method.

\subsection{Anomaly-free theory and smeared BCFT: Boundary states outside of the boundary of lattice models}
Here, we study a class of anomaly-free $Z_{N}$ theories in the sense of the LSM and simple current orbifolding. The anomaly-freeness results in the absence of the obstruction for the gauging or group extension\cite{Tachikawa:2017gyf,Bhardwaj:2017xup}. Let us start from the $Z_{N}$ symmetric bosonic theory with $Z_{N}$ simple current $J$ with the following modular invariant,
\begin{equation} 
Z=\sum_{i,p}\chi_{i,p}\overline{\chi}_{\overline{i},-p}+\sum_{a}|\chi_{a}|^{2}
\label{boson_modular_invariant}
\end{equation}
where $i$ is a label of modular noninvariant sectors, $a$ is that of $Z_{N}$ invariant sectors and $p$ is the $Z_{N}$ parity or $Z_{N}$ particle number couting taking $p=0,1,..,N-1$. The $Z_{N}$ simple current is a chiral primary field acting like a generator of $Z_{N}$ symmetry, such as $J\times\phi_{\alpha}=\phi_{\beta}$ where $\phi$ is a primary field. This kind of modular invariants appears in $\{ SU(N)_{1}\}^{k}$ WZW models or general $Z_{N}$ symmetric models with $N$ prime number (Similar arguments applies to more general cases by considering the prime decomposition of the number $N$ corresponding to subgroup of $Z_{N}$, and the author will present more general results elsewhere).
We have followed the notation in the author's previous works\cite{Fukusumi_2022_c,Fukusumi:2024cnl}. The anomaly-free condition\cite{Schellekens:1990ys,Gato-Rivera:1991bqv,Gato-Rivera:1991bcq,Kreuzer:1993tf} can be summarized as follows,
\begin{itemize}
\item{($N$ integer) $h_{J^{k}}=$integer for all $k$ and $N$,}
\item{($N$ even case) $h_{J^{k}}=$ integer for even $k$ and $h_{J^{k}}=$ half-integer for odd $k$,}
\end{itemize}
We propose that the above condition combined with gauging operation reproduces the anomaly-free condition in \cite{Kong:2019cuu}. The same conditions appeared in \cite{Fuchs:1997af,Schweigert:2000ix} and more recently in \cite{Cappelli:2010jv,Viola:2011tto,Schoutens:2015uia,Hung:2015hfa,Kikuchi:2022ipr,Fukusumi_2022_c}.

We introduce the following $Z_{N}$ charge of chiral primary fields,
\begin{equation}
Q_{J}(\alpha)=h_{J}+h_{\alpha}-h_{J\times \alpha}
\end{equation}
where $h$ is the conformal dimension of the corresponding field.
This is called $Z_{N}$ simple current charge or chiral t'Hooft anomaly in the high energy physics and mathematical physics communities and LSM charge in the condensed matter community. The anomaly-free theory satisfies the following mutual charge condition,
\begin{equation}
Q_{J}(J^{k})=0, \ \text{for}\ k=0,...,N-1
\end{equation}
In other words, the $Z_{N}$ particles do not have a twist as $Z_{N}$ symmetry generators each other. This is analogous to (quark) confinement in quark-hadron physics and the anomaly-free $Z_{N}$ simple current can be thought of as an object analogous to a hadron\cite{Schellekens:1990ys,Gato-Rivera:1991bqv,Gato-Rivera:1991bcq,Kreuzer:1993tf,Fuchs:1996rq,Fuchs:1996dd,Fukusumi_2022_c}. This absence of obstruction for condensation corresponds to gappability of spin chains known as Haldane conjecture or LSM theorem\cite{Lieb:1961fr,Schultz:1964fv,Haldane:1981zza,Affleck:1986pq,MOshikawa_1992,PhysRevB.34.6372,Cabra:1998vw,Lecheminant:2015iga,Cho:2017fgz,Yao:2018kel,Wamer:2019oge,Lecheminant:2024apm}.

For the latter discussions, we note the set of objects $\{ \Phi \}$ in spherical fusion category (SFC) $\mathbf{B}$ or nonchiral fusion rule of the bulk primary operators \cite{Rida:1999xu,Rida:1999ru,Balaska:2002qw,Balaska:2003xx} as follows,
\begin{align}
&\Phi_{i,p,-p}, 
\label{SFC_i}
\\
&\Phi_{a,0}.
\label{SFC_a}
\end{align}
The algebraic data of them is determined by the ring isomorphism to the chiral algebra, $\{ \Phi\} \leftrightarrow \{ \phi\}$ where the algebraic structure of $\{ \Phi\}$ or the MTC is determined by the Verlinde formula\cite{Verlinde:1988sn,Cardy:1989ir}. For later use, we introduce the notation for the $Z_{N}$ invariant sector of CCFT as $\theta_{a,0}=\phi_{a}$.
The objects in a SFC correspond to the bulk primary fields of the bosonic modular invariant as can be confirmed from the lower indices in Eq. \eqref{boson_modular_invariant} and Eq. \eqref{SFC_i} and Eq.\eqref{SFC_a}. The simple current extension of this algebra  denoted as $\mathbf{F}$ is,
\begin{align}
&\Phi_{i,p,\overline{p}}, \\
&\Phi_{a,p}.
\end{align}
where we have applied chiral parity shift operation $\mathcal{Q}_{J}: \Phi\rightarrow J\Phi$ to the SFC and extended the $Z_{N}$ invariant sectors labeled by $a$. Because the simple current is not included in the original SFC, the simple current extension or $Z_{N}$ gauging changes the system fundamentally. This is different from orbifolding which requires identifications of fields\cite{Dijkgraaf:1989hb}.
By applying anyon condensation or topological holography to this extended theory, one can obtain the algebraic data of the corresponding SymTFT $\mathbf{S}$. The algebraic data of SymTFT $\mathbf{S}=\{ \Psi \}$ can be identified as\cite{Fukusumi:2024cnl}, 
\begin{align}
\Psi_{i,p}= \frac{\sum_{p'}\Phi_{i,p'+p,-p}}{N}, \\
\Psi_{a,p}=\frac{\Phi_{a,p}}{\sqrt{N}}.
\end{align}
The CFT/TQFT correspondence results in the following correspondence (or ring isomorphism) between the SymTFT and the $Z_{N}$ extended CCFT $\{ \phi \}$,
\begin{align}
\phi_{i,p}&=\Psi_{i,p}, \\
\phi_{a,p}&=\Psi_{a,p}.
\end{align}
Because of the extension, the index $a$ in Eq. \eqref{boson_modular_invariant} is generalized to $(a,p)$. More mathematically, the simple current extension of a bosonic CCFT is an extension of $Z_{N}$ symmetric MTC to $Z_{N}$ graded chiral algebra (In algebraic level, the theory $\{ \phi \}$ is well-defined, but it is not sure this is in the scope of the existing category theory. This might correspond to the premodular fusion category (PMF) or superfusion category, but this identification is not conclusive. We thank Ingo Runkel and Steven Simon for reminding us the significance of this aspect.).

We also note the resultant chiral and antichiral decompositions (by Deligne product \cite{Deligne1990}),
\begin{align}
\Phi_{i,p,\overline{p}}&=\phi_{i,p}\overline{\phi}_{\overline{i},\overline{p}}, \\
\Phi_{a,p}&= \sum\frac{\phi_{a,p+p'} \overline{\phi}_{\overline{a},-p}}{\sqrt{N}}
\end{align}
with the identification of the following partition function characterized by charge,
\begin{equation}
Z_{Q}=\sum_{Q_{J}(i)=Q }|\sum_{p}\chi_{i,p}|^{2}+N\sum_{Q_{J}(a)=Q }|\chi_{a}|^{2}
\label{universal_seed}
\end{equation}

In the above discussions, the construction of CCFT has been shown. Hence we proceed to the construction of BCFT which is determined by the algebraic data of CCFT under the open-closed duuality\cite{Cardy:1989ir}.
By applying the gauging operation for the bosonic BCFT as in \cite{Fukusumi:2020irh,Fukusumi:2021zme,Weizmann}, one can obtain the set of $Z_{N}$ symmetric Cardy states as,
\begin{align} 
|i\rangle\rangle_{Z_{N}}&=\sum_{p} |i,p\rangle\rangle ,\\
|a\rangle\rangle_{Z_{N}}&=|a\rangle\rangle
\end{align}
The right-hand side is the bosonic Cardy state. One can check their $Z_{N}$ invariance by the general relation $\mathcal{Q}_{\alpha}|\beta \rangle \rangle=\sum_{\gamma}n^{\gamma}_{\alpha\beta}|\gamma\rangle\rangle $ in \cite{Graham:2003nc} and by taking $\alpha$ as the $Z_{N}$ simple current $J$. We also note the identification $\theta_{a,0}=\sum_{p}\phi_{a,p}/\sqrt{N}$ coming from the $Z_{N}$ extension of the theory. Hence under the fusion rule and Cardy's condition, one can obtain the following alternative expression,
\begin{equation}
|a\rangle\rangle_{Z_{N}}=\sum_{p}  \frac{1}{\sqrt{N}}|a,p\rangle\rangle
\end{equation}
The Cardy states $|a,p\rangle\rangle$ correspond to the $Z_{N}$ generalization of Majorana zero modes studied in the context of fermionic string theories and BCFTs\cite{Runkel:2020zgg,Smith:2021luc,Weizmann,Fukusumi:2021zme}. We note that the original algebraic structure of bosonic Cardy states can be recovered from this identification.

\begin{figure}[htbp]
\begin{center}
\includegraphics[width=0.5\textwidth]{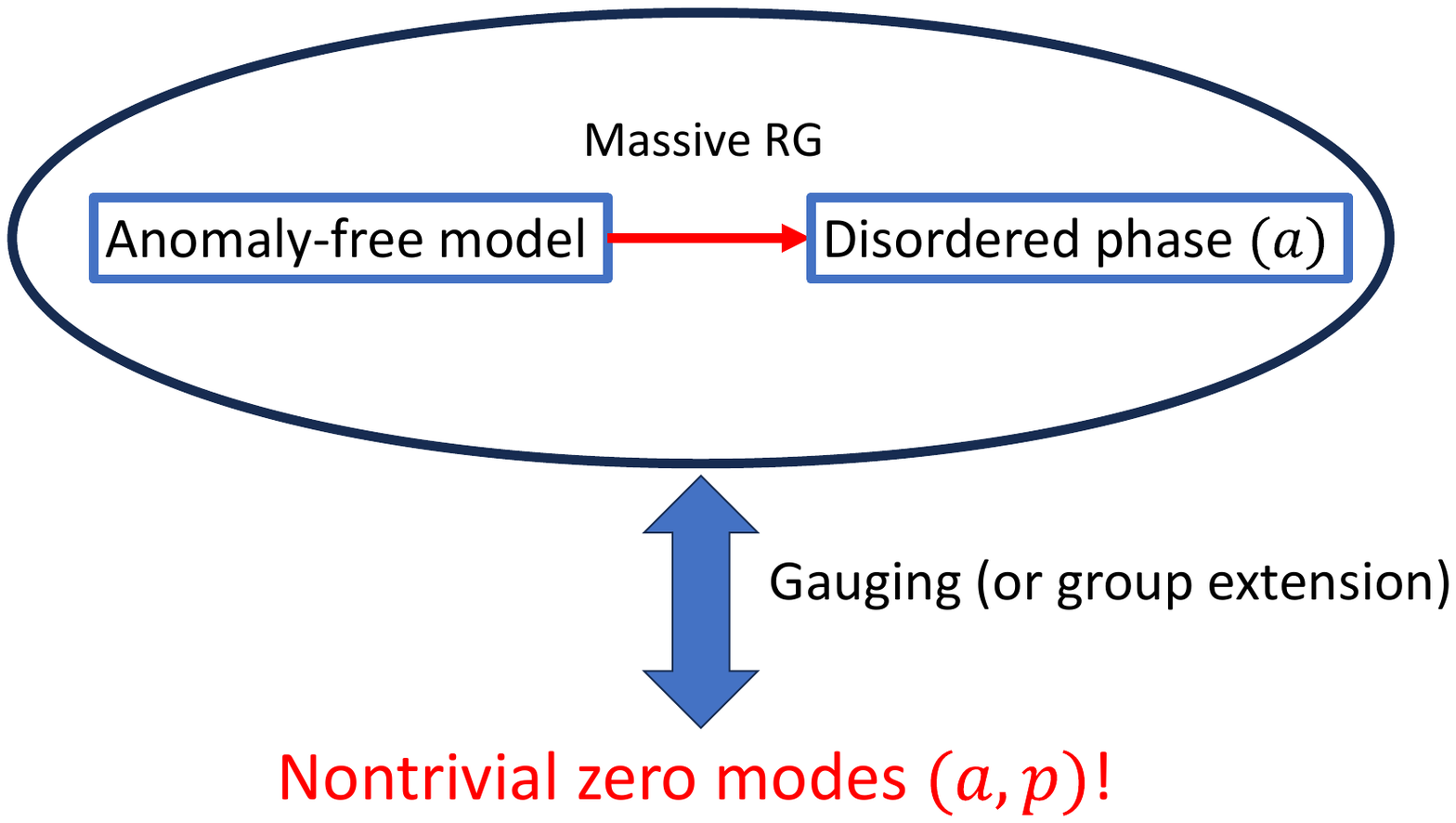}
\caption{Existence of a disordered state and its implication to an anomaly-free model. If we assume an $Z_{N}$ symmetric anomaly-free theory flows to a nondegenerate disordered phase labelled by $a$, the anomaly freeness imposes the existence of zero modes $(a,p)$. These zero modes only appear when applying the $Z_{N}$ extension to the resultant gapped system and are absent in the original spin or bosonic system.}
\label{anomaly_free_zero_modes}
\end{center}
\end{figure}

These boundary conditions labeled by $i$ and $a$ correspond to the boundary conditions of a quantum spin chain or its gauged version because the fusion rule produces only positive amplitudes. Hence this representation is called nonnegative-integer-matrix (NIM) representation. However, in the $Z_{N}$ gauged theories, there can exist twisted boundary conditions,
\begin{align} 
|i\rangle\rangle_{Z_{N}}^{(l)}&=\sum_{p} \omega^{lp} |i,p\rangle\rangle, 
\label{smeared_SSB_1}
\\
|a\rangle\rangle_{Z_{N}}^{(l)}&=\sum_{p}  \frac{\omega^{lp}}{\sqrt{N}}|a,p\rangle\rangle
\label{smeared_SSB_2}
\end{align}
where $\omega=e^{\frac{2\pi i}{N}}$ is the $N$-th root of unity and $l=0,..,N-1$ is the label of twist. One can implement this twist by applying the $Z_{N}$ parity operator. Hence, the set of boundary states is labeled by $(i,l)$ and $(a,l)$. The most significant property of this representation is,
\begin{equation}
\langle \langle \alpha|^{(l)}_{Z_{N}}|\alpha'\rangle \rangle^{(l')}_{Z_{N}}=0,
\end{equation}
for $l\neq l'$. The fusion rule is not in the scope of NIM representations, and unusual vanishing of the amplitude appears in these states. Hence, the total theory without fixing $l=0$ cannot apply to boundary critical phenomena of the $1+1$ dimensional quantum lattice model. However, if one interprets the boundary states as smeared BCFT, this vanishing and modified matrix representation can be allowed because this is a common structure of the inner product of (gapped) quantum states. 

Based on the arguments above, we propose,
\begin{itemize}
\item{The classification of SSB phase in bosonic representation corresponds to the classification of $\{ |i\rangle\rangle^{(l)}\}$. This also corresponds to the $1+1$ dimensional topological phase of gauged models\cite{Kitaev:2001kla,2014AnPhy.351.1026G}.}
\item{The classification of disordered states (typically disordered phase) in bosonic representation corresponds to the classification of $\{ |a\rangle\rangle^{(0)}\}$. The corresponding quantum phase generates zero modes  $\{ |a\rangle\rangle^{(l)} \}$ under gauging (Fig. \ref{anomaly_free_zero_modes}).}
\end{itemize}
In this sense, the smeared BCFTs corresponding to the gapped quantum states are more general than existing BCFTs.

To see the benefit of this representation, we study the application of the Majorana-Ising smeared BCFT corresponding to the gapped phase. In the $Z_{2}$ SSB phase, the corresponding smeared Cardy states are
\begin{equation}
|I \rangle \rangle \pm |\psi \rangle \rangle
\end{equation}
They correspond to the many-body cat state with spin up and down in the bosonic representation and Majorana zero mode in the fermionic representation. By changing the basis more convenient for the bosonic representation, one can recover the original bosonic BCFT breaking $Z_{2}$ symmetry spontaneously,
\begin{equation}
|I \rangle \rangle, \ |\psi \rangle \rangle
\end{equation}
They exchange each other by the application $\mathcal{Q}_{\psi}$ corresponding to the $Z_{2}$ spin flip operation.
Next, we discuss the aspect of Kramers-Wannier duality in this case. Both the above bosonic Cardy states transform to the following $Z_{2}$ symmetric states,
 \begin{equation}
|\sigma\rangle \rangle,
\end{equation}
under the application $\mathcal{Q}_{\sigma}$ corresponding to the KW dual transformation\cite{Frohlich:2004ef,Frohlich:2006ch,Ares:2020uwy}.
Hence, the duality is a two-to-one mapping and noninvertible in bosonic representation.

However, the fermionic T duality is invertible. By applying the corresponding operation $\mathcal{Q}_{e}$ or $\mathcal{Q}_{m}$ to $|I\rangle \pm |\psi\rangle$, one obtains,
\begin{equation}
|e \rangle \rangle \pm |m \rangle \rangle
\end{equation}
By modifying the normalization, these two states result in the following forms,
\begin{equation}
|\sigma\rangle \rangle,|\mu \rangle \rangle.
\end{equation}
Hence, one can confirm that the fermionic theory is the correct extension of the bosonic one, by taking the subset $\{ I, \psi, \sigma\}$.

The states $|e \rangle \rangle \pm |m \rangle \rangle$ correspond to the fermionic zero modes in the Ramond sector and also to the $Z_{2}$ extension of the disordered phase with a unique ground state. Moreover, under assuming some additional symmetries from the lattice, the disordered phase corresponds to the Haldane phase. There exists a general correspondence between the charged fermionic zero modes and the Haldane phase of the spin $S$ antiferromagnetic Heisenberg chain with $S$ odd integer by studying the corresponding  $SU(2)_{2S}$ Wess-Zumino-Witten (WZW) model\cite{Wess:1971yu,Witten:1983ar}. The $Z_{2}$ symmetric operator labeled by $a$ becomes $Z_{2}$ charged objects, only when the conformal dimension of the $Z_{2}$ simple current $J$ is half-integer. By considering the corresponding smeared Cardy states labeled by $a$, this explains the existing classification in \cite{Furuya:2015coa,MOshikawa_1992,Pollmann:2009ryx,Pollmann:2009mhk} in a simple but reasonable way. The conformal dimension of the $Z_{2}$ simple current in the corresponding WZW model becomes a half-integer only when the spin $S$ is an odd integer. The mapping between the Haldane and fermionic topological phase in \cite{Zirnbauer:2020nab} corresponds to this observation. Moreover, this argument seems to be connected to the degeneracies of the entanglement spectrum because of its close connection to zero modes $(a,p)$. Hence, one can obtain the following phenomenology,
\begin{equation}
\begin{split}
&\{ \text{ Anomaly classification of group invariant fields} (a) \}\\
&\leftrightarrow \{ \text{Classification of disordered (or SPT) state} \} \\
&\leftrightarrow\{ \text{Spin statistics of} \ J\}
\end{split}
\end{equation}
In other words, \emph{the underlying bosonic or fermionic statistics \cite{Pauli:1940zz} of $Z_{N}$ particle $J$  govern the SPT classifications}. This classification will be useful for the $Z_{2N}$ models because these models admit both fermionic and bosonic statistics. Similar periodicity can be seen in \cite{Zhou:2021ulc} and a possible generalization to massless RG flows can be seen in \cite{Herviou:2023unm}.

\subsection{Anomalous theory and massive flow: spontaneous symmetry breaking versus breaking of extended symmetry}

Next, let us study the anomalous $Z_{N}$ model. In anomalous theory, the $Z_{N}$ invariant sector $a$ does not appear \emph{without extension}(The absence of the $Z_{N}$ invariant sector in an anomalous theory comes from the well-definedness of the $Z_{N}$ charge.) However, still one can implement the $Z_{N}$ gauged Cardy states as,
\begin{equation}
|i\rangle\rangle_{Z_{N}}^{(l)}=\sum_{p} \omega^{lp} |i,p\rangle\rangle
\end{equation}
For example in the topological phase of $Z_{N}$ parafermion chain or SSB phase in the corresponding spin chain, the $N$-fold degenerated groundstates can be identified as $\sum_{p} \omega^{lp} |\psi_{p}\rangle\rangle$ or $|\psi_{p}\rangle\rangle$ where $\psi_{p}$ is the parafermionic field. (For the readers interested in the stability of the edge parafermionic zero modes, we note \cite{Fendley:2012vv,Jermyn_2014}.)

Because the theory is anomalous, the parity index $p$ carries charge. Hence, the original bosonic smeared Cardy state $ |i,p\rangle\rangle$ has different charges depending on both $i$ and $p$. This is a modified LSM-type argument using smeared Cardy states. One can see related arguments in \cite{Han:2017hdv,Numasawa:2017crf,Yao:2018kel,Kikuchi:2019ytf,Li:2022drc,Kikuchi:2022ipr}.

\begin{figure}[htbp]
\begin{center}
\includegraphics[width=0.5\textwidth]{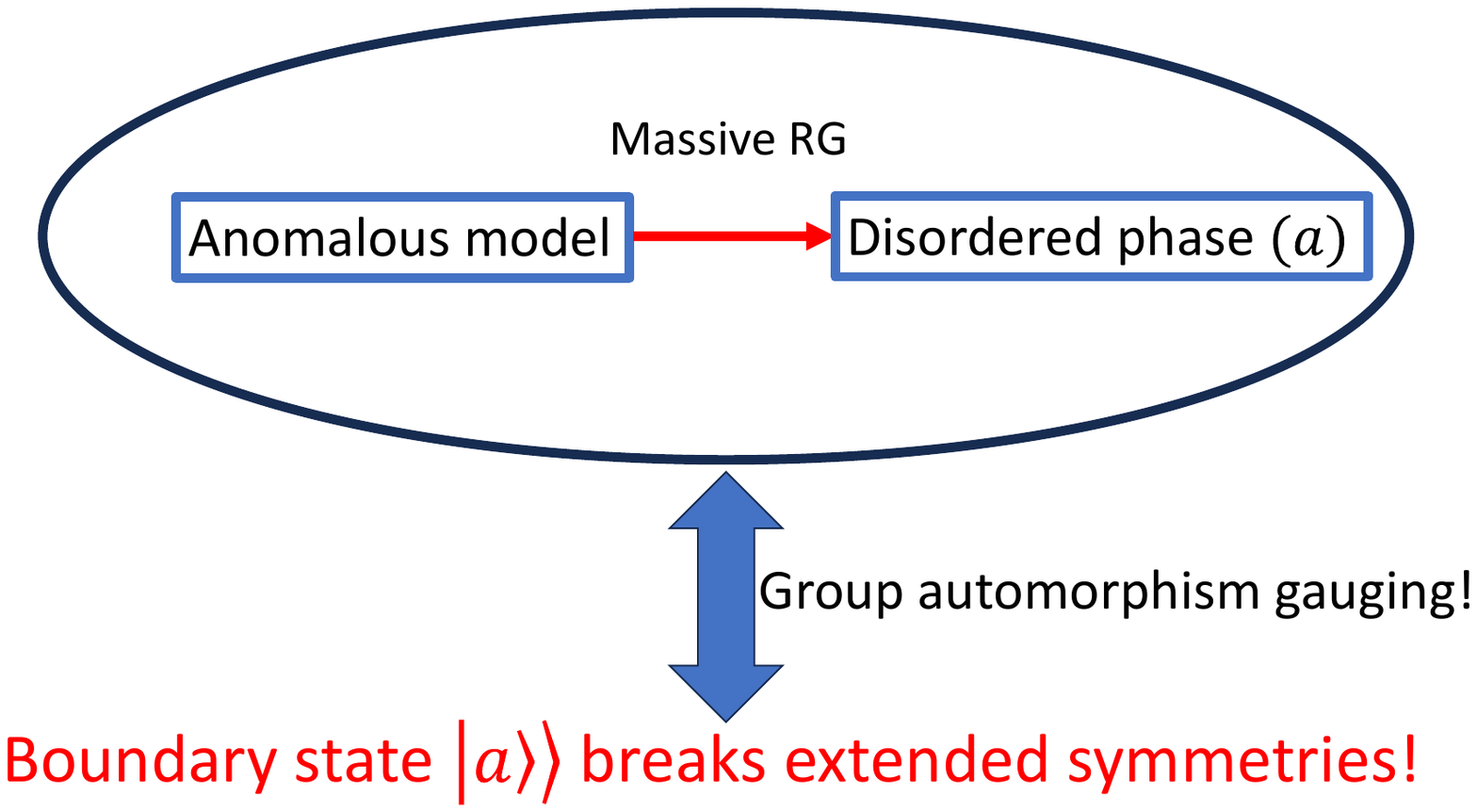}
\caption{Existence of disordered phase and its implication in an anomalous theory. If an anomalous $Z_{N}$ symmetric system flows to a nondegenerate gapped phase labelled by $a$, this indicates the existence of symmetry-breaking boundary condition in the original CFT. In other words, the information of massive RGs will provide nontrivial data of new or unfamiliar BCFTs.}
\label{anomalous_symmetry_breaking}
\end{center}
\end{figure}

In the parafermionic models, the disordered phase is more nontrivial from this view. For simplicity, let us concentrate our attention on the three-state Potts model. If one admits the conjecture by Cardy for this case, one needs to implement $Z_{N}$ invariant boundary states. Moreover, the resultant Cardy state should be related to the SSB phase under the Kramers-Wannier transformation. Hence, by applying the analogy to the Ising model, one has to introduce the free boundary condition $|\sigma\rangle \rangle$ as a $Z_{3}$ invariant boundary state. In literature, the corresponding operator $\sigma$ is a field in the $Z_{2}$ orbifolded theory of the three-state Potts model known as the minimal model $M(6,7)$\cite{Affleck:1998nq}, and this is outside of the primary fields of the original three-state Potts CFT. Whereas the original theory is an anomalous $Z_{3}$ model, the orbifolded theory is an anomaly-free $Z_{2}$ theory. In other words, the extended models appear as a disordered phase. However, in general, the extended object should break extended symmetry such as W-symmetry or Kac-Moody Lie group symmetry. These extended boundary conditions are called symmetry-breaking branes or symmetry-breaking boundary conditions in high energy physics \cite{Fuchs:1999zi,Birke:1999ik,Fuchs:1999xn,Quella:2002ct,Ishikawa:2002wx,Ishikawa:2005ea} and new boundary conditions in condensed matter or statistical mechanical physics\cite{Affleck:1998nq,Behrend:2000us,Iino:2020ipa} (we note the textbook \cite{Recknagel:2013uja} for reference). Based on this argument, one can understand the appearance of a disordered gapped phase in an anomalous system as a signal of the existence of an extended model (Fig. \ref{anomalous_symmetry_breaking}). Hence such an anomalous theory can be classified by the extended model (or anomalous theory should be considered as a consequence of the orbifolding of an anomaly-free model).

\section{Massless flow and its implications in topological order}
\label{section_massless_flow_TQFT}

In the previous section, we concentrated on the RG flow from CFTs to the gapped phase in $1+1$ dimensions. Under CFT/TQFT, this is equivalent to the bulk property of TOs in $2+1$ dimensions. However, in RGs from a CFT, there exists another situation where the system flows to other CFTs, called massless RG flow\cite{Zamolodchikov:1989hfa}. In this section, we revisit the problem by studying the tensor product of the CFTs connected by RG flow while preserving $Z_{N}$ symmetry. The construction of modular partition functions of coupled topologically ordered systems is proposed from a modern view and they contain a new series of modular partition functions that correspond to FQHE with nonchiral anyons. We show two types of theories by the different anomaly classification mechanisms, anomaly cancellation for the usual coupled (or layered) model and anomaly matching for the nonchiral theory.  The Halperin states\cite{Halperin:1983zz} or the FQHE in HLL are typical examples of the former and latter cases respectively. Moreover, by using the folding trick, we demonstrate that even the known partition can be interpreted as a nontrivial domain wall changing the chirality of anyons.

First, we note the general form of the modular $T^{2}$ (or $T$) invariant parition function describing the resultant coupled model,

\begin{equation}
Z_{Q}=\sum_{Q_{J_{\text{tot}}}(i_{\text{tot}})=Q }|\sum_{p}\Xi_{i_{\text{tot}},p}|^{2}+N\sum_{Q_{J_{\text{tot}}}(a_{\text{tot}})=Q }|\Xi_{a_{\text{tot}}}|^{2}
\label{universal_partition_function}
\end{equation}
where $\Xi$ is the character corresponding to the two types of theories, anomaly cancellation models and anomaly matching models which we will specify, and $J_{\text{tot}}$ is the corresponding (possibly non-chiral) $Z_{N}$ field. The prefactor $N$ in the equation corresponds to the zero modes. We propose this partition function provides a general classification of the gapless edge modes of topological order and this can be interpreted as a detailed expression of the proposal by Kong and Zheng \cite{Kong:2019cuu}. We remark that the significance of the studies on integer simple current again\cite{Schellekens:1990ys,Gato-Rivera:1991bqv,Gato-Rivera:1991bcq,Kreuzer:1993tf,Fuchs:1996rq,Fuchs:1996dd} because this new form is nothing but a generalization of Eq. \eqref{universal_seed}. 

The above partition function corresponds to $Z_{N}$ extension of the SFC $\mathbf{B}_{\text{UV}}\otimes \mathbf{B}_{\text{IR}}$ induced by the $Z_{N}$ group generator $J_{\text{tot}}$. Hence we denote the fusion algebra of this extended theory as $\mathbf{F}_{\text{UV}\times \text{IR}}$ (Recently the structure corresponding to $\mathbf{F}_{\text{UV}\times \text{IR}}$ has been studied in \cite{Antinucci:2025uvj}, and this work contains useful calculations of current operators. However, the corresponding coset representations have already been studied in \cite{Gaiotto:2012np} and the resultant partition function in the present work is more general). In the pioneering work \cite{Gaiotto:2012np}, it has been conjectured that the BCFT of the extended theory $\mathbf{F}_{\text{UV}\times \text{IR}}$ in the present work produces the information of the massless RG between the original $\mathbf{B}_{\text{UV}}$ and $\mathbf{B}_{\text{UV}}$ theories (one can see the phenomenology in the figure 2 of \cite{Gaiotto:2012np} in a concise way). We also remind that the appearance of extended model for the classification of BCFTs is common in the studies of symmetry breaking boundary conditions\cite{Affleck:1998nq,Recknagel:2013uja}. We also note that this algebra $\mathbf{F}_{\text{UV}\times \text{IR}}$ is included in $\mathbf{F}_{\text{UV}}\otimes \mathbf{F}_{\text{IR}}$ but they are not equivalent. Depending on the class of the theories or the form of $J_{\text{tot}}$, the character $\Xi$ becomes chiral or nonchiral and this corresponds to the non-chiral anyon in \cite{Kong:2019cuu}. The total charge $Q_{J_{\text{tot}}}$ is the generalized $Z_{N}$ charge (or anomaly) defined by the conformal spin (or by conformal dimensions indirectly) of the operators. For simplicity, we concentrate our attention on the unitary models, but one can generalize the arguments to nonunitary models, by replacing the definition of charge as monodromy charge\cite{Gannon:2003de} (For the readers interested in the application of nonunitary CFTs to TOs, we note works on gapless FQHEs \cite{PhysRevLett.60.956,PhysRevB.75.075317,Bernevig2008PropertiesON,Bernevig_2008,Davenport:2012fcs,Yang_2021}).

The anomaly-free conditions in our convention are relatively simple,
\begin{itemize}
\item{($N$ integer) $s_{J_{\text{tot}}^{k}}=$integer for all $k$ and $N$,}
\item{($N$ even case) $s_{J_{\text{tot}}^{k}}=$ integer for $k$: even and $s_{J_{\text{tot}}^{k}}=$ half-integer for $k$: odd,}
\end{itemize}
where $s$ is the conformal spin of the operators labeling $\Xi$. Under these conditions, the theory also satisfies the LSM anomaly-free condition,
\begin{equation}
Q_{J_{\text{tot}}}(J_{\text{tot}}^{k})=0
\end{equation}
which we will introduce. In the literature on category theories, the above anomaly-free condition may have been written in some abstract ways, but we have written it more directly.

This condition is a natural generalization of the integer spin simple current by Schellekens and Gato-Rivera \cite{Schellekens:1990ys,Gato-Rivera:1991bqv,Gato-Rivera:1991bcq,Kreuzer:1993tf,Fuchs:1996rq,Fuchs:1996dd} revisited recently in \cite{Hung:2015hfa,Fukusumi_2022,Fukusumi_2022_c,Fukusumi:2022xxe,Kikuchi:2022ipr,Fukusumi:2023psx,Fukusumi:2024cnl}. The corresponding discussion can be seen in the pioneering work by Kong and Zheng \cite{Kong:2019cuu} (and related recent work \cite{Kaidi:2021gbs}) by using category theory and particular cases have been studied in detail by Kikuchi\cite{Kikuchi:2022ipr,Kikuchi:2024cjd} in the context of massless RGs (also see recent work \cite{Benedetti:2024utz}). Our arguments below can be interpreted as a unification and generalization of the literature. Moreover, the algebraic data of simple current extension of CCFTs have been provided by the sections before and one can obtain the asymptotic behaviors of the corresponding wavefunctions systematically by combining the methods in \cite{Lu_2010,Schoutens:2015uia,Fukusumi:2023psx}. The algebraic data in this work will be useful also for further numerical research by using matrix-product states (MPS)\cite{Zaletel_2012,Estienne_2013,estienne2013fractionalquantumhallmatrix,Wu_2015,Vanhove:2018wlb,Chen:2022wvy,Ueda:2024jzj}.

\subsection{Massless flow as coupled topological orders}

For more specific discussions, we introduce two $Z_{N}$ symmetric models connected under an RG flow as,
\begin{align}
Z_{\text{UV}}&= \sum_{\alpha_{\text{UV}}} |\chi_{\alpha_{\text{UV}}}|^{2},\\
Z_{\text{IR}}&=\sum_{\alpha_{\text{IR}}} |\chi_{\alpha_{\text{IR}}}|^{2}.
\end{align}
where $\alpha$ is the label of primary fields of the bosonic theory represented as $(i,p)$ and $a$. One of the most fundamental points of the study of the RG domain wall is the correspondence between the conformal interface between UV and IR CFT and the BCFT of the product theory $\mathbf{F}_{\text{UV}\times \text{IR}}$\cite{Gaiotto:2012np}. 

Because $Z_{\text{UV}}$ and $Z_{\text{IR}}$ are modular invariant, one can implement the product $Z_{\text{UV}} Z_{\text{IR}}$ as a modular invariant. However, the partition function $Z_{\text{UV}} Z_{\text{IR}}$ is a trivial theory because all of the charge sectors in the UV and IR theories mix randomly. Hence, this theory itself does not produce a nontrivial conformal interface with 't Hooft anomaly matching condition straightforwardly. Here we introduce one type of anomaly freeness of the product theory, 
\begin{itemize}
\item{($N$: Integer) $h_{J^{k}_{\text{UV}}J^{k}_{\text{IR}}}=$ integer for all $k$ and $N$}
\item{($N$: Even integer) $h_{J^{k}_{\text{UV}}J^{k}_{\text{IR}}}$= integer for $k$: even and $h_{J^{k}_{\text{UV}}J^{k}_{\text{IR}}}$= half-integer for $k$: odd}
\end{itemize}
where $J$ is the UV and IR $Z_{N}$ simple current which is a chiral primary field acting like the group generator in the fusion rules as in the previous sections. We call this condition as anomaly cancellation condition. 
For the latter discussions, we remind the definition of $Z_{N}$ charge,
\begin{equation}
Q_{J}(\alpha)=h_{J}+h_{\alpha}-h_{J\times \alpha}
\end{equation}
where $\times$ represents the fusion product. Because the theory is anomaly-free, one can apply the gauging operation to the system. Depending on whether $h_{J^{k}_{\text{UV}}J^{k}_{\text{IR}}}$ is integer or half-integer the theory become modular $T$ or $T^{2}$ invariant.

Then, let us study the case where the UV $Z_{N}$ CFT is anomalous. In this setting, the IR CFT is also anomalous because of the anomaly-freeness of the coupled theory. The resultant IR theory should also be anomalous because of the anomaly-freeness of the product theory. Hence\, one can obtain the modular $S$ invariant,
\begin{equation}
Z_{0}=\sum_{i_{\text{UV}},i_{\text{IR}}} |\sum_{p}\chi_{i_{\text{UV}},p}\chi_{i_{\text{IR}},p}|^{2}
\end{equation}
where the total charge of each sector $Q_{\text{UV}\times \text{IR}}=Q_{\text{UV}}+Q_{\text{IR}}$ becomes zero. Hence, this is more convenient than the original $Z_{\text{UV}}Z_{\text{IR}}$ when studying symmetry (or charge) preserving domain wall. Similarly one can introduce the twisted partition function as
\begin{equation}
Z_{nQ_{1}}=\sum_{i_{\text{UV}},i_{\text{IR}}} |\sum_{p}\chi_{i_{\text{UV}},p+n}\chi_{i_{\text{IR}},p}|^{2}
\end{equation}
where we have taken a notation with $Q_{1}=Q_{J_{\text{UV}}}(J_{\text{UV}})$ and $(i,0)$ sector for the UV and IR theory is not charged.

The set of operators corresponding to the modular $S$ invariant are
\begin{equation}
\phi_{i_{\text{UV}},p}\phi_{i_{\text{IR}},p}\overline{\phi}_{\overline{i}_{\text{UV}},\overline{p}}\overline{\phi}_{\overline{i}_{\text{IR}},\overline{p}},
\end{equation}
.

When the UV theory is anomaly-free, the corresponding partition function is,

\begin{equation}
\begin{split}
Z_{Q}&=\sum_{i_{\text{UV}},i_{\text{IR}},p':Q_{\text{UV}\times \text{IR}}=Q} |\sum_{p}\chi_{i_{\text{UV}},p+p'}\chi_{i_{\text{IR}},p}|^{2} \\
&+\sum_{i_{\text{UV}},a_{\text{IR}}:Q_{\text{UV}\times \text{IR}}=Q} |\sum_{p}\chi_{i_{\text{UV}},p}\chi_{a_{\text{IR}}}|^{2} \\
&+\sum_{a_{\text{UV}},i_{\text{IR}}:Q_{\text{UV}\times \text{IR}}=Q} |\sum_{p}\chi_{a_{\text{UV}}}\chi_{i_{\text{IR}},p}|^{2} \\
&+N\sum_{a_{\text{UV}},a_{\text{IR}}:Q_{\text{UV}\times \text{IR}}=Q} |\chi_{a_{\text{UV}},}\chi_{a_{\text{IR}}}|^{2},
\end{split}
\end{equation}

The above analysis gives criteria for studying the massless RG flow of $Z_{N}$ models such as $SU(N)_{k}$ WZW model. For the $SU(N)_{k}$ WZW models, this produces the RG connectivity of $SU(N)_{k}$ and $SU(N)_{k'}$ models with periodicy $k+k' (\text{mod} N)$. We call this massless flow anomaly cancellation flow because of the anomaly cancellation condition of UV and IR theories. The resultant coupled model gives the anomaly-free theory and this can be understood as bilayer or multicomponent systems. In the anomaly cancellation case, the chiral integer current $J_{\text{tot}}=J_{\text{UV}}J_{\text{IR}}$ and the corresponding extension plays a fundamental role.

One expects another class of product theory corresponding to the RG connectivity between $SU(N)_{k}$ and $SU(N)_{k'}$ with $k-k' (\text{mod}.N)$. However, the corresponding product theory does not satisfy the same anomaly-free condition as in the anomaly cancellation cases when $N$ is greater than $3$. Hence, it is necessary to implement a different type of $Z_{N}$ gauging procedure to understand the product theory as an anomaly-free theory. The fundamental points of the existing anomaly free theory is $Q_{\text{UV}\times \text{IR}}(J)=0$, and modular $T$ and $T^{2}$ invariance. Hence we modify the definition of $Q_{\text{UV}\times \text{IR}}$ by introducing a new $Z_{N}$ symmetry generator.

For this purpose, we introduce the new nonchiral simple current as $J_{\text{tot}}=j\overline{j}=J_{\text{UV}}\overline{J}_{\text{IR}}$. The modular property is defined as the conformal spin of this object,
\begin{equation}
s_{j\overline{j}}=h_{J_{\text{UV}}}-\overline{h}_{\overline{J}_{\text{IR}}}
\end{equation}
and the condition for the bosonic spin of $j\overline{j}$ produces the periodicity  $k-k' (\text{mod}.N)$ in the $SU(N)_{k}$ WZW models. Because of the anomaly matching between UV and IR theories, we name the corresponding RG flow anomaly matching flow. More generally, the anomaly matching condition can be summarized as,

\begin{itemize}
\item{($N$ integer) $s_{(j\overline{j})^{k}}=$integer for all $k$ and $N$,}
\item{($N$ even case) $s_{{(j\overline{j})}^{k}}=$ integer for $k$: even and $s_{(j\overline{j})^{k}}=$ half-integer for $k$: odd,}
\end{itemize}

Related arguments for nonchiral theories can be seen in \cite{Fuji_2017,Sule:2013qla,Sohal_2020} and our argument is a natural extension of them. By the definition of the charge conjugation, one can obtain the anomaly-fee condition $Q_{j\overline{j}}(j\overline{j})=Q_{J_{\text{UV}}}(J_{\text{UV}})-Q_{\overline{J}_{\text{IR}}}(\overline{J}_{\text{IR}})=0$.

The modular partition function is,
\begin{equation}
Z_{Q}=\sum_{i_{\text{UV}},i_{\text{IR}}:Q_{\text{UV}\times \text{IR}}=Q} |\sum_{p}\chi_{i_{\text{UV}},p}\overline{\chi}_{\overline{i}_{\text{IR}},p}|^{2}
\end{equation}
for an anomalous UV theory and 
\begin{equation}
\begin{split}
Z_{Q}&=\sum_{i_{\text{UV}},i_{\text{IR}},p':Q_{\text{UV}\times \text{IR}}=Q} |\sum_{p}\chi_{i_{\text{UV}},p+p'}\overline{\chi}_{\overline{i}_{\text{IR}},p}|^{2} \\
&+\sum_{i_{\text{UV}},a_{\text{IR}}:Q_{\text{UV}\times \text{IR}}=Q} |\sum_{p}\chi_{i_{\text{UV}},p}\overline{\chi}_{\overline{a}_{\text{IR}}}|^{2} \\
&+\sum_{a_{\text{UV}},i_{\text{IR}}:Q_{\text{UV}\times \text{IR}}=Q} |\sum_{p}\chi_{a_{\text{UV}}}\overline{\chi}_{\overline{i}_{\text{IR}},p}|^{2} \\
&+N\sum_{a_{\text{UV}},a_{\text{IR}}:Q_{\text{UV}\times \text{IR}}=Q} |\chi_{a_{\text{UV}},}\overline{\chi}_{\overline{a}_{\text{IR}}}|^{2},
\end{split}
\end{equation}
for an anomaly-free UV theory where we have taken the summation for the $i$ and $a$ indices in both UV and IR theories. By taking the total charge $Q_{\text{UV}\times \text{IR}}$ to $0$, one obtains a modular invariant.

At this stage, the above arguments seem like formal calculations of modular invariants, but they exhibit nontrivial properties in the context of TOs. When interpreting the above partition function as that of a TO in cylinder geometry, this implies the existence of nonchiral anyons,
\begin{equation}
\phi_{\alpha_{\text{UV}}}\overline{\phi}_{\overline{\alpha}_{\text{IR}}}.
\end{equation}
with $Z_{N}$ electron operator $j\overline{j}$. 
The single-valuedness of the wavefunctions is satisfied by the anomaly-freeness of the operator  $j\overline{j}$.
Hence, the resultant wavefunctions in plane geometry can contain both contributions from holomorphic and antiholomorphic variables. To distinguish this theory from a usual CCFT appearing as edge modes of topological order, we call this \emph{one edge CFT} (OECFT). As far as we know, a similar mixing of the holomorphic and antiholomorphic pairs appears in the study of FQHEs with nonchiral anyons, such as antiPfaffian and HLL. Hence, we propose that \emph{FQHE with nonchiral anyon is described by OECFT corresponding to anomaly-matched RG flow.} Similar observations can be seen in \cite{Kaidi:2021gbs}, and our proposal can be considered as a generalization of their arguments \footnote{We thank Justin Kaidi for related discussions}.

\begin{figure}[htbp]
\begin{center}
\includegraphics[width=0.5\textwidth]{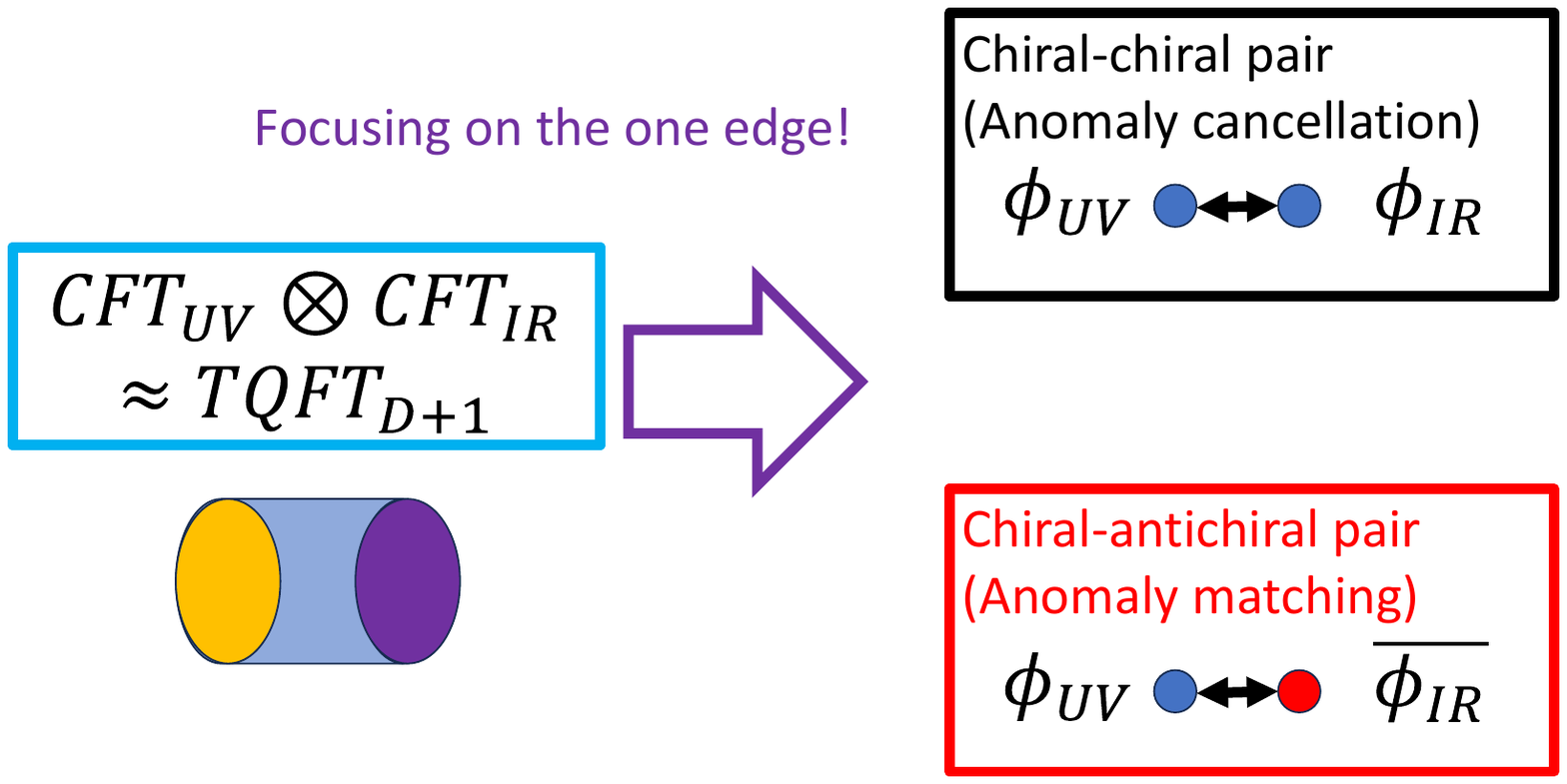}
\caption{OECFT constructed from the product of $Z_{N}$ extended theories. By interpreting the partition function as that of $CFT_{UV}\otimes CFT_{IR}$ and focusing on the chiral-chiral or chiral antichiral structure $\Xi$, one can obtain the pairing structures of UV and IR theories. The chiral-chiral pairs appear in the systems with anomaly cancellation, and the chiral-antichiral pairs appear in the systems with anomaly matching.}
\label{CFT_OECFT}
\end{center}
\end{figure}

Before moving to the next subsection, we comment on the generalization of our method to multicomponent systems. The above discussions can be easily generalized to the theories constructed from coupled three or more CFTs. For this purpose, let us assume the OECFT has $N_{(+)}$ CCFT and $N_{(-)}$ antichiral CFTs with simple current $\{ J_{n_{(+)}}\}_{n_{(+)}=1}^{N_{(+)}}$ and $\{J_{n_{(-)}}\}_{n_{(-)}=1}^{N_{(-)}}$. By redefining the $J_{\text{tot}}$ as $J_{\text{tot}}=\prod_{n_{(+)},n_{(-)}}J_{n_{(+)}}\overline{J}_{n_{(-)}}$ and applying the anomaly-free condition for the total conformal spin $s_{\text{tot}}=\sum_{n_{(+)}}h_{J_{n_{(+)}}}-\sum_{n_{(-)}}\overline{h}_{J_{n_{(-)}}}$, one can generalize the arguments in this section to this situation.

The resultant partition function for the anomaly-free case is unchanged and one can still obtain,
\begin{equation}
Z_{Q}=\sum_{Q_{J_{\text{tot}}}(i_{\text{tot}})=Q }|\sum_{p_{\text{tot}}}\Xi_{i_{\text{tot}},p_{\text{tot}}}|^{2}+N\sum_{Q_{J_{\text{tot}}}(a_{\text{tot}})=Q }|\Xi_{a_{\text{tot}}}|^{2}
\end{equation}
where $i_{\text{tot}}$, $p_{\text{tot}}$, $a_{\text{tot}}$ are the label of tensor product of the theories.

By studying the charge defined by $J_{\text{tot}}$ for the tensor product of OECFT $\otimes_{n_{(+)}}\mathbf{S}_{n_{(+)}}\otimes_{n_{(-)}}\overline{\mathbf{S}}_{n_{(-)}}$ or $Z_{N}$ extended fusion category $\otimes_{n_{(+)}}\mathbf{F}_{n_{(+)}}\otimes_{n_{(-)}}\overline{\mathbf{F}}_{n_{(-)}}$ , one can obtain the corresponding fusion rules. Moreover, by modifying the definition of the total charge defined by conformal dimension to the monodromy charge defined by the modular $S$ transformation, one can apply the same phenomenology to nonunitary cases. We note the form of monodromy charge as,
\begin{equation}
\begin{split}
Q_{J_{\text{tot}}}^{\text{mon}}( \ ) &=Q_{J_{\text{tot}}} ( \ )- Q_{J_{\text{tot}}}(\sigma_{\text{tot}}) \\
&=\sum_{n_{(+)}}Q_{J_{n_{(+)}}}(\ )-\sum_{n_{(-)}}Q_{\overline{J}_{n_{(-)}}}(\ )  \\
&-\sum_{n_{(+)}}Q_{J_{n_{(+)}}}(\sigma_{n_{(+)}})+\sum_{n_{(-)}}Q_{\overline{J}_{n_{(-)}}}(\overline{\sigma}_{n_{(-)}})
\end{split}
\end{equation}
where $\{ \sigma \}$ are the primary fields with the lowest conformal dimension corresponding to the effective vacuum of each theory and we replaced the total objects in our interests as blank by treating $Q_{J_{\text{tot}}}^{\text{mon}}$ as a function. We note literature on massless flow of coset models related to the multicomponent models\cite{LeClair:2001yp,Lecheminant:2002va,Georgiou:2017jfi,Sfetsos:2017sep,Georgiou:2018gpe,Borsato:2023dis} studied with a connection to $\lambda$ deformations, and older works on coset construction for nonunitary CFTs \cite{Schellekens:1989uf,Saleur:2003zm,Saleur:2001cw,Mathieu:2000hg}, and recent works on RG flows between nonunitary CFTs\cite{Nakayama:2024msv,Katsevich:2024jgq,Katsevich:2024sov,Delouche:2024tjf}. By applying the correspondence between the unitary and nonunitary CFTs or symplectic duality realized by a similarity transformation\cite{Guruswamy:1996rk,Hsieh:2022hgi}, further studies on models constructed by our method will be useful for the classification of TOs in non-hermitian or corresponding (nonlocal) Hermitian systems. We note recent works \cite{Lootens:2019xjv,Chen:2021obc,Xu:2021zzl} useful for this research direction.

\subsection{Massless flow as the interface of topological orders}

\begin{figure}[htbp]
\begin{center}
\includegraphics[width=0.5\textwidth]{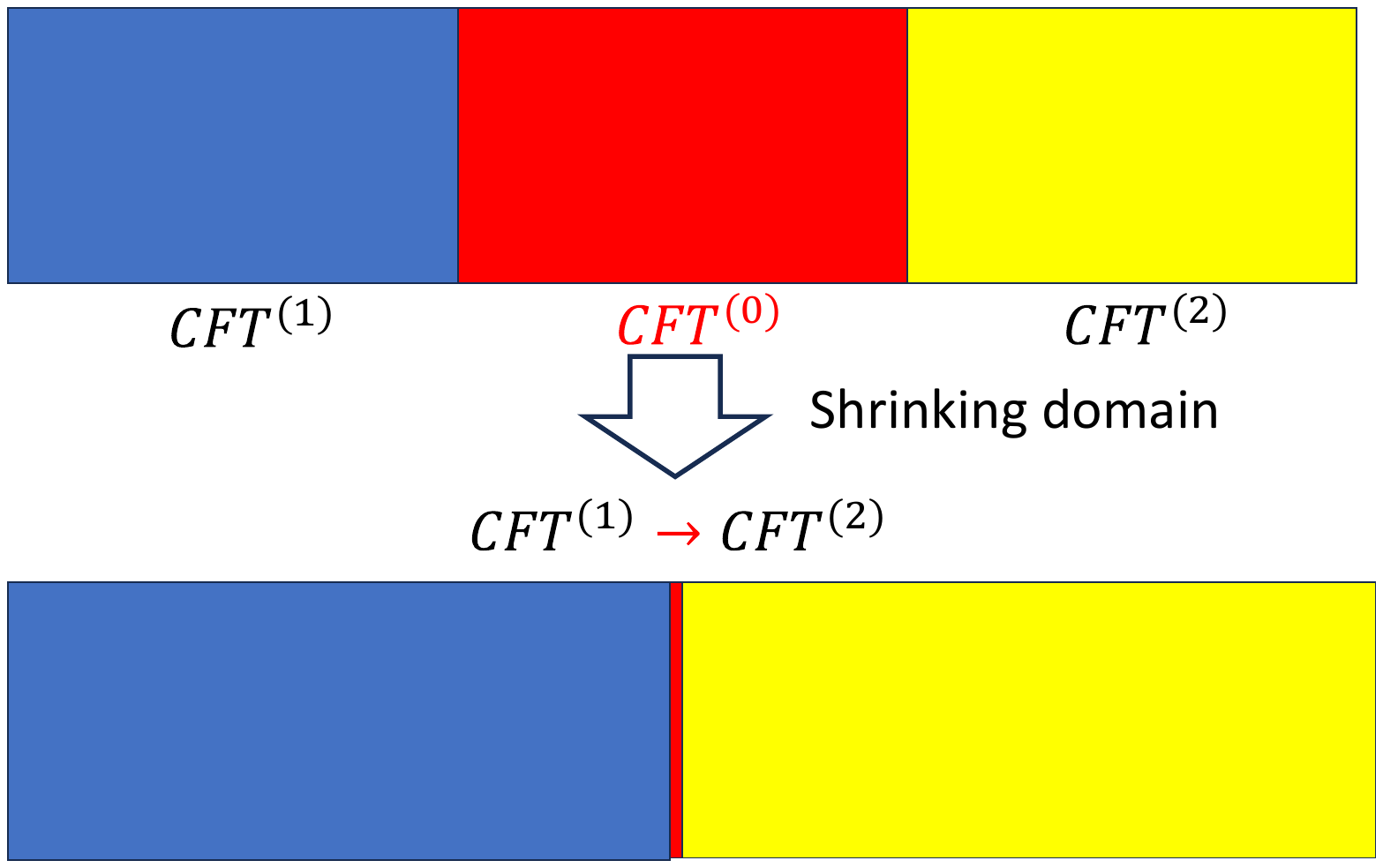}
\caption{Multiple domain walls and the consequence of shrinking. One can map anyons in one system to another (between $CFT^{(1)}$, $CFT^{(0)}$, and $CFT^{(2)}$ in the figure), and change the shape of the system consistently (resulting in $CFT^{(1)}\rightarrow CFT^{(2)}$). By changing the geometry of the settings, one can apply our analysis to multi-junction problems (To obtain analytical data, more respective calculations are necessary). Similar ideas can be seen commonly in literature\cite{Bachas:2007td,Huston:2022utd,Zhao:2023wtg,Barter:2018hjs,Bridgeman:2018jdv}.}
\label{domain_fusion}
\end{center}
\end{figure}

\begin{figure}[htbp]
\begin{center}
\includegraphics[width=0.5\textwidth]{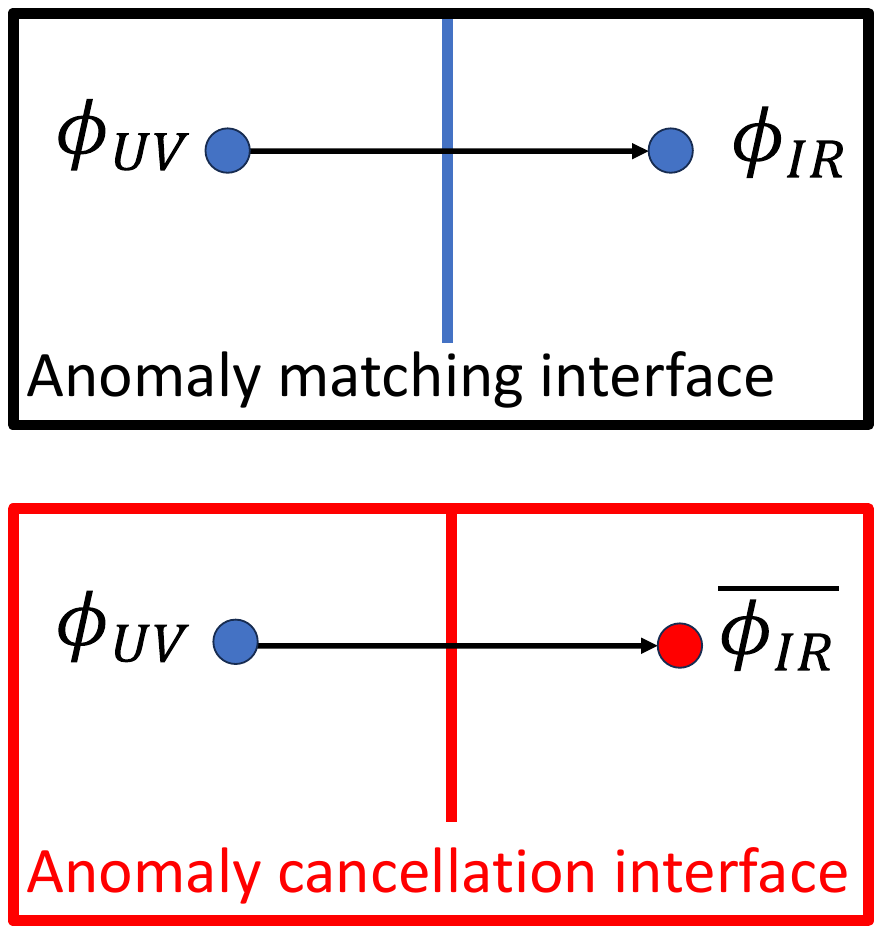}
\caption{Picture of the respective interface. After the folding trick\cite{Wong:1994np}, the simple current extended partition function implies the mapping relation between anyons in the theories. The role of chiral and antichiral pair has been conjugated compared with Fig.\ref{CFT_OECFT}.}
\label{anyon_transformation}
\end{center}
\end{figure}

In the previous subsection, we studied the products of theories connected by the RG domain wall. By applying the folding trick\cite{Wong:1994np} to this coupled model, it is possible to interpret the set of partition functions as a mapping of theories. Hence, there exists following mapping $D_{\text{RG}}$ corresponding to the modular invariant or the charge $0$ sector of the product of the theories $\{\mathbf{F}_{\text{UV}}\otimes \mathbf{F}_{\text{IR}}\}_{Q_{\text{tot}}=0}$,

\begin{equation} 
D_{\text{RG}}: \ \mathbf{F}_{\text{UV}} \rightarrow \overline{\mathbf{F}_{\text{IR}}},
\label{Mapping_RG}
\end{equation}
where $\mathbf{F}$ is the $Z_{N}$ extension of SFC corresponding to $Z_{N}$ extended bulk CFT. We note that it is possible to implement the domain walls that do not preserve the anomaly by applying the folding trick to the coupled model with the charged partition functions. It may be reasonable to call them $Z_{N}$ charged domain walls analogous to the domain walls studied in astrophysics\cite{Campanelli:2003cs,Kibble:1976sj,Campanelli:2004si,Campanelli:2005xy,Wang:2009ay,Belendryasova:2017wad}. We concentrate our attention on $Q_{\text{tot}}=0$ for simplicity, but by attaching charged particle (analogous to quark) to the domain wall $Q_{\text{tot}}=0$ or studying the structure of modular covariant $Q_{\text{tot}}\neq 0$, one can implement the charged domain wall. We also remark that one can generalize the setting to multi-domain walls and shrink them freely (Fig. \ref{domain_fusion}). Hence, by this construction, one can obtain the possible set of domain wall particles systematically. We also note recent studies on domain wall in condensed matter \cite{Huston:2022utd,Bao_2022,Zhao:2023wtg,Barter:2018hjs,Bridgeman:2018jdv,Li:2024crt}. In principle, one can apply our analysis to the corresponding quantum multi-junction problems (or Kondo problem)\cite{Affleck:1990by,Kane:1992xse,Kane:1992zza,Polchinski:1994my,Fendley_1995,Fendley:1998sq,Lesage:1998jky,Oshikawa:2005fh,Bachas:2007td,Roy:2023wer}. For the readers interested in the theoretical analysis of junction problem, we note two reviews\cite{Affleck:1995ge,Saleur:1998hq}.  

Roughly speaking, one can identify Eq. \eqref{Mapping_RG} providing a mapping relation between two topological orders in a cylinder geometry. Because the Witt equivalence of existing MTCs is too strong to apply for the analysis of the RG domain wall, we have introduced $Z_{N}$ extension of SFCs and defined its anomaly-free conditions as an equivalence relation. We also remind that the fundamental point connecting the phenomenology of interface and that of the product theory is the folding trick\cite{Wong:1994np}. Hence, the role of chiral and antichiral parts of either $UV$ or $IR$ theory should be swapped. Roughly speaking, one can understand each object appearing in the set of modular partition functions on the basis of $\{\mathbf{F}_{\text{UV}}\otimes \mathbf{F}_{\text{IR}}\}_{Q_{\text{tot}}=0}$ as the following mapping,

\begin{equation}
\begin{split}
&\{ \Phi_{\alpha_{\text{UV}}}\Phi_{\alpha_{\text{IR}}}\in \mathbf{F}_{\text{UV}}\otimes \mathbf{F}_{\text{IR}}\}_{Q_{\text{tot}}=0}
\\
&\Rightarrow \{ D_{\text{RG}}: \Phi_{\alpha_{\text{UV}}} \rightarrow \overline{\Phi}_{\overline{\alpha}_{\text{IR}}} \}_{Q_{\text{UV}=Q_{\text{IR}}}}
\end{split}
\end{equation}
where $\overline{\alpha}$ is the complex conjugation of $\alpha$ and we droped the parity indeces for simplicity. However, this only implies the condition for the charge and parity, and more detailed data are necessary for pratical calculations. Hence, the following discussion, we assume that the construction of the RG domain wall has been established by more respective techniques as in \cite{Crnkovic:1989ug,Quella:2006de,Gaiotto:2012np,Brunner:2015vva,Kimura:2015nka,Stanishkov:2016pvi,Stanishkov:2016rgv,Klos:2019axh,Klos:2021gab,Poghosyan:2022mfw,Poghosyan:2022ecv,Zeev:2022cnv,Fukusumi:2023vjm,Konechny:2023xvo,Poghosyan:2023brb,Kikuchi:2022ipr,Kikuchi:2024cjd}  and we concentrate our attention on the properties obtained from the charge and parity structures. Surprisingly, only from these structures, one can obtain nontrivial domain walls changing the chirality of anyons.

By the above observations and taking bulk semion algebra \cite{Fukusumi:2024cnl} (or by topological holography\cite{Moradi:2022lqp}), one can construct action of $D_{\text{RG}}$ to SymTFT because the SymTFT can be interpreted as a subalgebra of the SFC. 

\begin{equation}
\{ D_{\text{RG}}: \Phi_{\alpha_{\text{UV}}} \rightarrow \Phi_{\overline{\alpha}_{\text{IR}}} \}\Rightarrow \{ D_{\text{RG}}: \Psi_{\alpha_{\text{UV}}} \rightarrow \Psi_{\overline{\alpha}_{\text{IR}}} \}
\end{equation}
The right arrow $\Rightarrow$ is obtained by bulk semionization or sandwich construction. To check consistency between this formalism and the domain wall Verlinde formula in \cite{Zhao:2023wtg} (or similar generalization \cite{Choi:2024tri}) is an interesting research direction. For simplicity, we study the models with anomalous UV and IR theories, but the same phenomenology should hold for the models with the anomaly free UV and IR theories.

By assuming the anomaly preservation $\{ Q_{\text{tot}}=0\}$ under the application of RG domain walls, one can restrict the form of the mapping as,
\begin{equation}
D_{RG}: \{ \phi_{\alpha_{\text{UV}},p} \}\rightarrow \{\phi_{\alpha_{\text{IR}},p}\} 
\end{equation}
with $Q_{J_{\text{UV}}}(\alpha_{\text{UV}})=Q_{J_{\text{IR}}}(\alpha_{\text{IR}})$ for the anonamly matching flow. The corresponding matrix representation is,
\begin{equation}
D_{\text{RG}}(\phi_{\alpha_{\text{UV}},p})=\sum_{\alpha'_{\text{IR}}:Q_{J_{\text{UV}}}(\alpha_{\text{UV}})=Q_{J_{\text{IR}}}(\alpha_{\text{IR}})}D^{\alpha'_{\text{IR}}}_{\alpha_{\text{UV}}}\phi_{\alpha'_{\text{IR}},p}
\end{equation} 
with the respective data of the matrix $D$.

By using the same technique, one can obtain the corresponding mapping for the anomaly cancellation case as,
\begin{equation}
D_{RG}: \{\phi_{\alpha_{\text{UV}},p}\} \rightarrow \{\overline{\phi}_{\overline{\alpha}_{\text{IR}},p}\} 
\end{equation}
with $Q_{J_{\text{UV}}}(\alpha_{\text{UV}})+Q_{\overline{J}_{\text{IR}}}(\overline{\alpha}_{\text{IR}})=0$.
The corresponding matrix representation is,
\begin{equation}
D_{\text{RG}}(\phi_{\alpha_{\text{UV}},p})=\sum_{\overline{\alpha'}_{\text{IR}}:Q_{J_{\text{UV}}}(\alpha_{\text{UV}})+Q_{\overline{J}_{\text{IR}}}(\overline{\alpha}_{\text{IR}})=0}D^{\overline{\alpha'}_{\text{IR}}}_{\alpha_{\text{UV}}}\overline{\phi}_{\overline{\alpha'}_{\text{IR}},p}
\end{equation}

 In both anomaly-matching and anomaly-cancellation cases, the $Z_{N}$ parity is preserved under the mapping when the UV theory is anomalous. Interestingly, in the anomaly cancellation flow, the role of chiral and antichiral fields is swapped under the application of the RG domain wall. 

One can think of a UV theory as a bulk TO and an IR theory as edge modes and vice versa by shrinking the region of one theory. In other words, in a realistic setting, the bulk-edge correspondence should be relaxed up to the RG connectivity and this is consistent with the nondecreasing of $g$-value\cite{Green:2007wr,Fukusumi:2020irh,Fukusumi:2023vjm} enabling emergence of edge physics.  Based on this view, one can discuss the emergent properties or stabilities of bulk and edge modes by comparing their central charges. This can explain the existing controversies on the thermal Hall conductance\cite{Mross_2018,Wang2017TopologicalOF} because a CFT describing the bulk wavefunction cannot uniquely determine the edge theories without additional assumptions. Related arguments can be also seen in \cite{Stone:2012ud,Son:2015xqa,Barkeshli:2015afa,Wan2015StripedQH,Yang_2017}. More respective assumptions such as symmetry protection are worth further investigation, but this may require more detailed data on the respective settings.

The above argument can be easily generalized to the multicomponent model. For this purpose, let us consider the situation where the two multicomponent TOs are connected in an anomaly-free way. We note the respective multicomponent $Z_{N}$ object as 
\begin{equation}
J^{(i)}=\prod_{n_{(+)}^{(i)}}^{N_{(+)}^{(i)}} J_{n_{(+)}^{(i)}}\prod_{n_{(-)}^{(i)}}^{N_{(-)}^{(i)}} \overline{J}_{n_{(-)}^{(i)}}
\end{equation}
where $(i)$ is the label of two systems and $\pm$ is labeling chiral and anithiral components. Hence, by applying the folding trick\cite{Wong:1994np}, the two theories are Witt equivalent when $J_{\text{tot}}=J^{(1)}\overline{J}^{(2)}$ satisfy the anomaly-free condition.

Before going to the concluding remark, we comment on the $Z_{N}$ charged domain wall in our setting. If one considers a $Z_{N}$ charged domain wall with $Q_{\text{tot}}\neq 0$, the $Z_{N}$ charge of the anyons is not preserved through the domain wall. Hence it is necessary to introduce some additional degree of freedom at the domain wall to recover the charge preservation. Compared with the original anomaly-free $Z_{N}$ current $J_{\text{tot}}$ analogous to hadron, this charged particle can be interpreted as a quark. This mechanism of cancellation of charge and the resultant appearance of nontrivial particles are known as anomaly-inflow and has been studied widely in both high energy physics and condensed matter theory community\cite{Callan:1984sa,Stone:2012ud}. One can summarize the phenomenology as follows,

\begin{equation}
\{\text{ Charged domain wall}\}\leftrightarrow \{\text{ Domain wall quark}\}
\end{equation}
For example, when attaching two Majorana CFTs with $Z_{2}$ charged domain wall, one can see domain wall semions because of the fermionic T duality\cite{Gliozzi:1976qd,Ginsparg:1988ui,Fukusumi:2022xxe}.  We note \cite{Ma:2021xqf,Ma:2022pnk,Ma:2024ivd} as references studying this case in details. This case can be regarded as a quark analog of domain wall fermion\cite{Kaplan:1992bt,Kaplan:1992sg,Narayanan:1993zzh,Narayanan:1993ss,Shamir:1993zy,Furman:1994ky,Vranas:2000tz}.

\section{Conclusion}
\label{conclusion}

\begin{figure}[htbp]
\begin{center}
\includegraphics[width=0.5\textwidth]{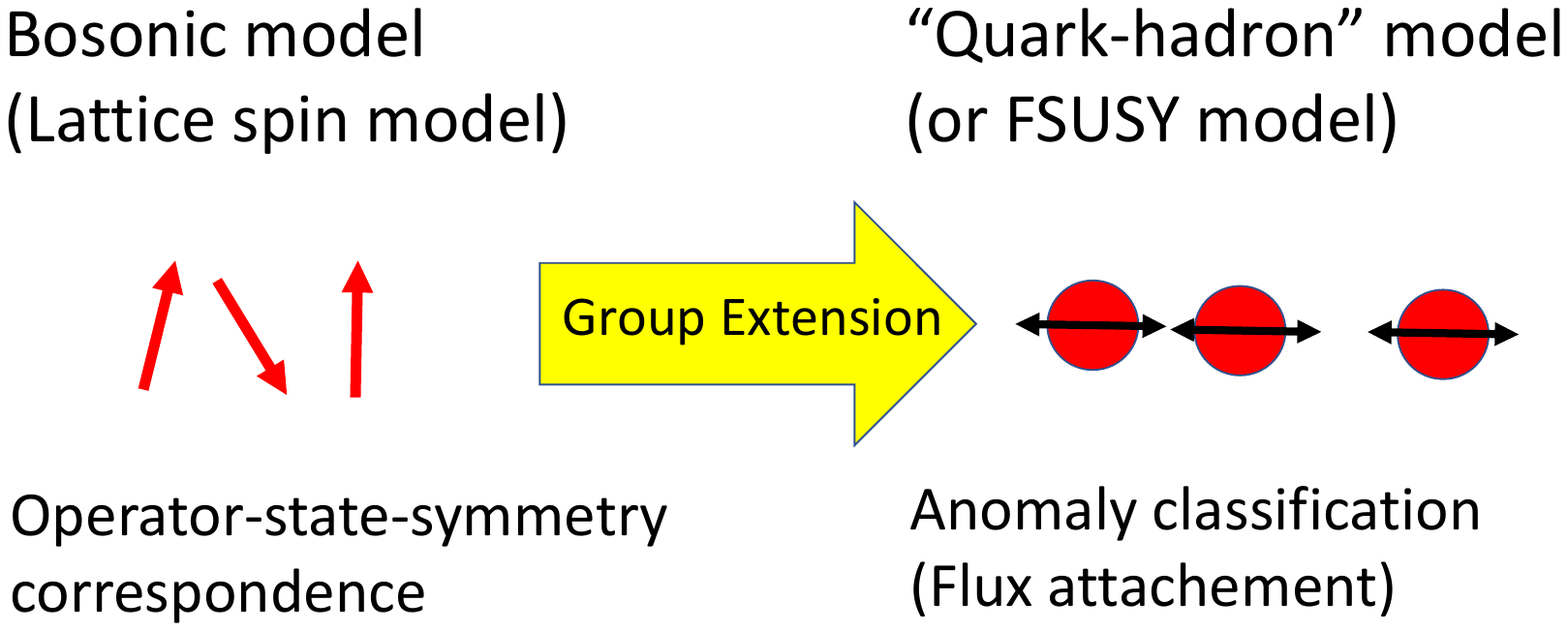}
\caption{Respective benefit of the bosonic representations and group extended representations. In a bosonic model, a correspondence between operators constructing correlation functions, quantum states, and (generalized) symmetry holds. Hence, one can achieve various calculations, by using vertex-operator-algebra or category theory and more. In a group-extended model analogous to a hadron model (or fractional supersymmetric model), the theory contains information on anomaly classifications determined by the conformal dimensions of the simple current. Moreover, one can attach gauge field easily. However, because the zero modes contain the nontrivial entanglement, calculations of the correlation functions become difficult.}
\label{boson_hadron}
\end{center}
\end{figure}

In this work, we have studied the general construction and classification of TOs based on the analysis of the integer spin simple current introduced by Schellekens and Gato-Rivera and generalized by Kong and Zheng. By combining the resultant anomaly analysis and recent studies on the massive and massless RG, we have obtained a general understanding of bulk and edge properties of $2+1$ dimensional TOs. We have constructed a series of unified bulk CFT, CCFT, and BCFT with a series of modular partition functions summarized as Eq. \eqref{universal_partition_function}. Moreover, the nonchiral modular invariants contain new series. The application of CFT/TQFT and the relation between RG flows in CFTs and Witt equivalent anyon condensation in TQFTs become more evident. We expect that our discussions will be useful for further studies by using numerical methods and the phenomenological explanation for the experimental settings.

For readers interested in the recent development of theoretical studies on TO or FQHE, we summarize the notions relevant in this manuscript. 
\begin{itemize}
\item{The bulk topological degeneracies are governed by (smeared) Cardy states}
\item{Multicomponent or multilayer TOs are constructed by anomaly cancellation mechanism}
\item{TOs with nonchiral anyons are obtained by considering anomaly matching mechanism.}
\item{FQHEs can be classified by anomaly matching through the interface. This implies the existence of the nontrivial dualities between FQHEs with chiral anyon and those with nonchial anyon. Moreover, the existence of a domain wall quark is implied from the charged domain wall.}
\end{itemize}
The first point is governed by the anyon condensation from the extended SFC or topological holography and the other points are governed by Witt equivalence combined with gauging and orbifolding.

As a concluding remark, we note the respective benefits of the bosonic and fermionic (or group extended) representations. Bosonic models have been extensively studied because this representation usually provides systematic calculations of correlation functions and related physical observables. In a modern understanding, the correspondence between quantum states, correlators, and (generalized) symmetry seems to be a fundamental source of this benefit. On the other hand, the group-extended model corresponds to a particle (or quark-hadron) picture. Hence, one can attach gauge fields and apply the ideas of condensations more easily (Fig. \ref{boson_hadron}). However, as we have observed in this work, the treatment of the zero modes and their entanglement structure requires careful treatment. We also remark on the difficulty of attaching gauge fields to bosonic systems\cite{Faddeev:1976pg,Alvarez-Gaume:1986nqf,Bloete:1986qm,Smilga:1994hc,Smilga:1996dn,Kitazawa_1997,Ikeda:2002wh,Thorngren:2018bhj,Yao:2019bub,Fukusumi:2021qwa}. (It may be surprising for the readers, but the naive coupling of bosonic fields and $U(1)$ gauge field cannot reproduce the LSM theorem and the spectrum of twisted models\cite{Alcaraz:1987zr,Alcaraz_1989}.) Because of the theoretical difficulties of the gauging operations, studies on the corresponding lattice or microscopic models are fundamentally important respectively. Inversely, the field-theoretical understanding of such models is also important for more general understanding of quantum phases.

\section{Acknowledgement}
We thank Rui-Dong Zhu for introducing the work \cite{Gaiotto:2012np} to us about six years ago and Yuji Tachikawa and Yunqin Zheng for a fruitful collaboration closely related to new BCFTs in this manuscript. We also thank Urei Miura, Shumpei Iino, Kohki Kawabata, Tokiro Numasawa, Steven Simon, Masahiko G. Yamada, and Yuan Yao for the interesting discussions. We thank Bo Yang and Guangyue Ji for the related collaborations and for notifying us of the recent progress on the studies of the HLL in FQHEs and for teaching us the basic notions in FQHE through their lectures and tutorials. We also thank Hao Gang for notifying the work\cite{Cogburn:2023xzw} for us. We acknowledge the support from NCTS and CTC-NTHU.

\appendix

\section{Insertion of topological defect and noninvertible chiral gauging}
\label{defect_symmetry_algebra}
In this section, we revisit the algebraic calculation of the formulation of topological defects and symmetry from a modern view. We provide an algebraic formalism to study the application of defects as chiral or antichiral fields in the SFC or its extension. Similar arguments can be seen in existing literature\cite{Petkova:2000ip}, but we pay attention to the chiral or antichiral structure. These objects correspond to the insertion of defects, and they are related to symmetries $\mathcal{Q}_{\alpha}$ in the main texts by exchanging the role of time and space by the modular $S$ transformation. 

First, let us concentrate our attention on the chiral objects $\phi$ in a single theory.
Roughly speaking, the multiplication of these objects to bulk fields $\{ \Phi\}$ changes the theory drastically because of the nontrivial conformal spin of the chiral objects. To implement the operation in a concrete form, it is necessary to define the algebraic relation containing both $\{ \Phi\}$ and $\{ \phi\}$. 

We introduce the chiral defect $\{ \mathcal{D}\}$ which are isomorphic to $\{ \phi\}$, to distinguish the chiral and antichiral decomposition of $\{ \Phi\} \sim \{ \phi\} \times \{ \overline{\phi}\}$. In other words, they satisfy the following relation,
\begin{equation}
\mathcal{D}_{\alpha} \times \mathcal{D}_{\alpha'} =\sum_{\gamma}n^{\gamma}_{\alpha\alpha'}  \mathcal{D}_{\gamma}
\end{equation}
where $n$ is the fusion matrix of $\{ \phi \}$ and is determined by the ring isomorihism between $\{ \phi\}$ and Sym TFT $\{\Psi\}$\cite{Fukusumi:2024cnl}. In bosonic theories, the fusion matrix is also determined by the Verlinde formula \cite{Cardy:1989ir,Verlinde:1988sn}.

In this representation, one can define the algebra of $\{ \Phi\} \otimes \{\phi\}$ by taking the respective tensor product,
\begin{equation}
\mathcal{D}_{\alpha} \Phi_{\beta}  \times \mathcal{D}_{\alpha'}\Phi_{\beta'} =\sum_{\gamma}n^{\gamma}_{\alpha\alpha'}\mathcal{D}_{\gamma} \sum_{\beta}N^{\delta}_{\beta\beta'} \Phi_{\delta}
\end{equation}
Hence, one can obtain the fusion rule of this nonlocal quantum field theory straightforwardly. 

Next, to construct the partition function, we introduce the following contraction of the chiral and antichiral decomposition of $\Phi$ as,
\begin{equation}
\mathcal{D}_{\alpha} \circ (\phi_{\beta} \overline{\phi}_{\overline{\beta}})=(\phi_{\alpha}\times \phi_{\beta}) \overline{\phi}_{\overline{\beta}}=\sum_{\gamma}n^{\gamma}_{\alpha,\beta}\phi_{\gamma}\overline{\phi}_{\overline{\beta}}
\end{equation} 
However, we stress that the resultant object produced by the contraction cannot necessarily imply the complete classification in a realistic setting. A famous example is the disorder field in the Ising model which can be constructed by multiplying the chiral Majorana fermion by the order operator. We explain this subtly more.

The Ising fusion rule leads to the following relation,
\begin{equation}
\mathcal{D}_{\psi} \circ \sigma \overline{\sigma}=\sigma \overline{\sigma},
\end{equation} 
However, it is known that the $\mathcal{D}_{\psi} \sigma \overline{\sigma}$ corresponds to the disorder operator. This has been established in the study of fermionic models and this counting is known as fermionic zero modes. Hence, the interpretation of objects after the contraction needs further careful discussion, whereas they provide the correct partition functions.

In the above discussions, one can see the correspondence between $\phi$ and $\mathcal{D}$ evidently. We stress that the formal chiral and antichiral decomposition of the bulk fields $\Phi$ is not a tensor product. This is called the Deligne product \cite{Deligne1990} verified by ``abstract nonsense" terminologically, but still, the formal decomposition is correct up to the chiral and antichiral decomposition of the partition function. Hence, one can trust the discussion on the amplitude in Section. \ref{CFT_symmetry}. As a summary, we note the relevant set of objects and identifications,
\begin{align}
&\mathcal{D}_{\alpha}\sim \phi_{\alpha}, \\
&\mathcal{D}_{\alpha}\Phi_{\beta}\\
&\mathcal{D}_{\alpha}\circ \phi_{\beta}= \phi_{\alpha}\times \phi_{\beta}. 
\end{align}
The algebraic structure of $\{ \mathcal{D}_{\alpha}\} \otimes \{ \Phi\}$ under the identification induced by the contraction $\circ$ is an interesting problem. This is an example of gauging operation by a generalized symmetry labeled by $\alpha$ in modern terminology.

The set $\{\mathcal{D}_{\alpha}\}$ corresponds to the chiral twist following modern literature on Fradkin-Kadanoff transformations\cite{Yao:2020dqx}. Hence, one can implement the antichiral twist under the following identification,
\begin{equation}
\overline{\mathcal{D}}_{\overline{\alpha}}\sim \overline{\phi}_{\overline{\alpha}}
\end{equation}

In successive discussions, we introduce the modular $S$ transformation to study the symmetry $\{\mathcal{Q}\}$. 
Here, we assume that the modular $S$ transformation does not change the total Hilbert space. Hence, the modular $S$ transformation is a mapping $S: \{\phi\}\times \{ \overline{\phi}\} \rightarrow \{ \Phi \}$. This can be achieved by constructing the modular $S$ matrix for the chiral and antichiral fields and applying the projection operator to the original Hilbert space. In other words, the construction of the original theory is fundamentally important. This choice of projection may become nontrivial when assuming the existence of extended chiral objects. For example, if one studies an orbifolded model, such as the $D$ series model, the object in the $A$ series model appears as such. Moreover, in the recent studies on Majorana and the parafermionic model, this extension of the Verlinde line also appears. Hence, the representation $S_{\{\Phi^{\text{ext}}\}}$ with the lower index specifying the extended theory is more appropriate in this situation. Because the modular $S$ transformation exchanges the role of space and time, one can obtain the following mutual mapping,
\begin{equation}
S:\mathcal{Q}_{\alpha}\leftrightarrow \mathcal{D}_{\alpha}
\end{equation}
Hence, by applying the property $S^{2}\sim 1$ of modular transformation, one can obtain the following expression,
\begin{equation}
\mathcal Q_{\alpha} \Phi_{\beta}=\mathcal{S}(\mathcal{D}_{\alpha}\circ \Phi_{\beta}).
\end{equation}
To distinguish its applications to the chiral characters, we denote the modular transformation as $\mathcal{S}$.
Whereas the expression $\mathcal{D}$ produces the extended theory, the application $\mathcal{Q}$ does not provide any notrivial objects without introducing $ \mathcal{S}_{\{\Phi\}^{\text{ext}}}$. 
The objects on the righthand side are completely determined by the algebraic relations. Hence, one can apply an arbitrary number of defects and symmetries. 

Here we note duality in the Ising CFT and Majorana CFT. By applying duality symmetry corresponding to order operator $\sigma$ with conformal dimension $1/16$ in the theory to the order operator, it seems to produce
\begin{equation}
\mathcal{Q}_{\sigma} \sigma \overline{\sigma}=0
\end{equation}
Because of the vanishing of the element $S_{\sigma}^{\sigma}$.

However, in the Majorana CFT, the application of the corresponding object to the same object becomes a disorder field and one obtains $\mathcal{Q}_{\sigma} \sigma_{\text{Bulk}}=\mu_{\text{Bulk}}$. In the Majorana CFT, one treats both the disorder operator and order operator as Ramond fields with the same conformal dimensions. Hence intertwining between the fields under the application of the symmetry happens\cite{Petkova:2000ip}. This view will be useful for further study on the construction and understanding of the order and disorder fields. One can see related discussions in \cite{Gaiotto:2012np,Kikuchi:2024cjd}.

\subsection{Gauging anomalous symmetry at one edge conformal field theories}
In this subsection, we note the result for the gauging of anomalous symmetry in the bulk OECFT. The method is the same as in the existing works\cite{Yao:2020dqx,Kawabata:2024hzx,Chen:2023jht,Fukusumi_2022,Fukusumi_2022_c}, so we only note the results,

\begin{equation}
Z_{Q}=\sum_{Q_{J_{\text{tot}}}(i_{\text{tot}},p)=Q }\Xi_{i_{\text{tot}},p}\left( \sum_{\overline{p}_{\text{tot}}} \overline{\Xi}_{\overline{i}_{\text{tot}},\overline{p}_{\text{tot}}}\right),
\end{equation} 
where we have used the same notation in the main text and $J_{\text{tot}}$ does not satisfy the anomaly-free condition.
Because of the anomaly, the definition of the charge contains the $Z_{N}$ parity indices. By replacing $\Xi\leftrightarrow \overline{\Xi}$ one can obtain the other gauging coming from the other edge in TQFT. Interestingly, whereas the total algebraic objects in the full theory are the same for these two theories, their sets of partition functions or sectors classified by their $Z_{N}$ charge are different. This difference prevents one from mapping the theories to the corresponding two-dimensional statistical models twisted by $Z_{N}$ Verlinde lines, whereas there can exist the corresponding one-dimensional quantum lattice models\cite{Yao:2020dqx,Fukusumi_2022,Fukusumi_2022_c}.

\section{Realization of group extension corresponding to anomaly matching in lattice model}
\label{Lattice_realization}
In this section, we propose the realization of the new nonchiral gauging operation implied from anomaly matching by coupling two quantum chains. The corresponding arguments for anomaly cancellation in the context of integer spin simple current can be seen in \cite{Fukusumi_2022_c} and \cite{Lahtinen_2021,Baver:1996kf} for related lattice models (We also note recent paper\cite{Wang:2023pkp} on paraparticles. Their models seem to correspond to our model but the precise relationship between our model and theirs is an open problem). For simplicity, let us concentrate our attention on the parafermion chain and corresponding $Z_{N}$ symmetric CFT\cite{Fateev:1985mm,Qiu:1987qm}. We mainly follow the approach in \cite{Fradkin:1980th,Yao:2020dqx,Fukusumi_2022_c}. Related discussions for parafermionic models without the coupling can be seen in \cite{Li_2015,Kawabata:2024hzx,Chen:2023jht}. The precise relationship between our approach and existing ladder or multicomponent $Z_{N}$ models \cite{Lahtinen_2021,Wouters_2022,Cao:2024qjj} is a future problem.

In a parafermion chain, one can interpret the chiral and antichiral $Z_{N}$ gauging operations in the $Z_{N}$ CFT as two different Fradkin-Kadanoff transformations. These two transformations correspond to the choice of disorder operator and produce two different CFTs because the theory is anomalous. To distinguish the two Fradkin-Kadanoff transformations, we label them chiral and antichiral Fradkin-Kadanoff transformations respectively.

Let us introduce the paraspin variable $\sigma_{l}$ and $\tau_{l}$ at site $l$ with the following algebraic relations,
\begin{equation}
\sigma^{N}_{l}=\tau^{N}_{l}=1, \ \sigma_{l}^{\dagger}=\sigma_{l}^{-1}, \ \tau_{l}^{\dagger}=\tau_{l}^{-1},  \ \sigma_{l}\tau_{l}=\omega \tau_{l}\sigma_{l},  
\end{equation}
where $\omega=\text{exp}(2 \pi i /N)$ as in the main text. Then the chiral parafermionization is defined as follows,
\begin{align}
\gamma_{2l-1}&=\sigma_{l}\mu_{l}  \\
\gamma_{2l}&=\omega^{(N-1)/2}\sigma_{l}\mu_{l}
\end{align}
where $\gamma$ is the chiral parafermionic variables satisfying the following relations,
\begin{equation}
\gamma_{l}\gamma_{l'}=\omega^{\text{sgn}(l-l')}\gamma_{l'}\gamma_{l}.
\end{equation}
One can also obtain the antichiral version by applying the space inversion. Because of the anomaly, the particle statistics changes to 
\begin{equation}
\gamma'_{l}\gamma'_{l'}=\omega^{-\text{sgn}(l-l')}\gamma'_{l'}\gamma'_{l}
\end{equation}
after the antichiral Fradkin-Kadanoff transformation where $\gamma'$ is the parafermionic variable after the transformation.

\begin{figure}[htbp]
\begin{center}
\includegraphics[width=0.5\textwidth]{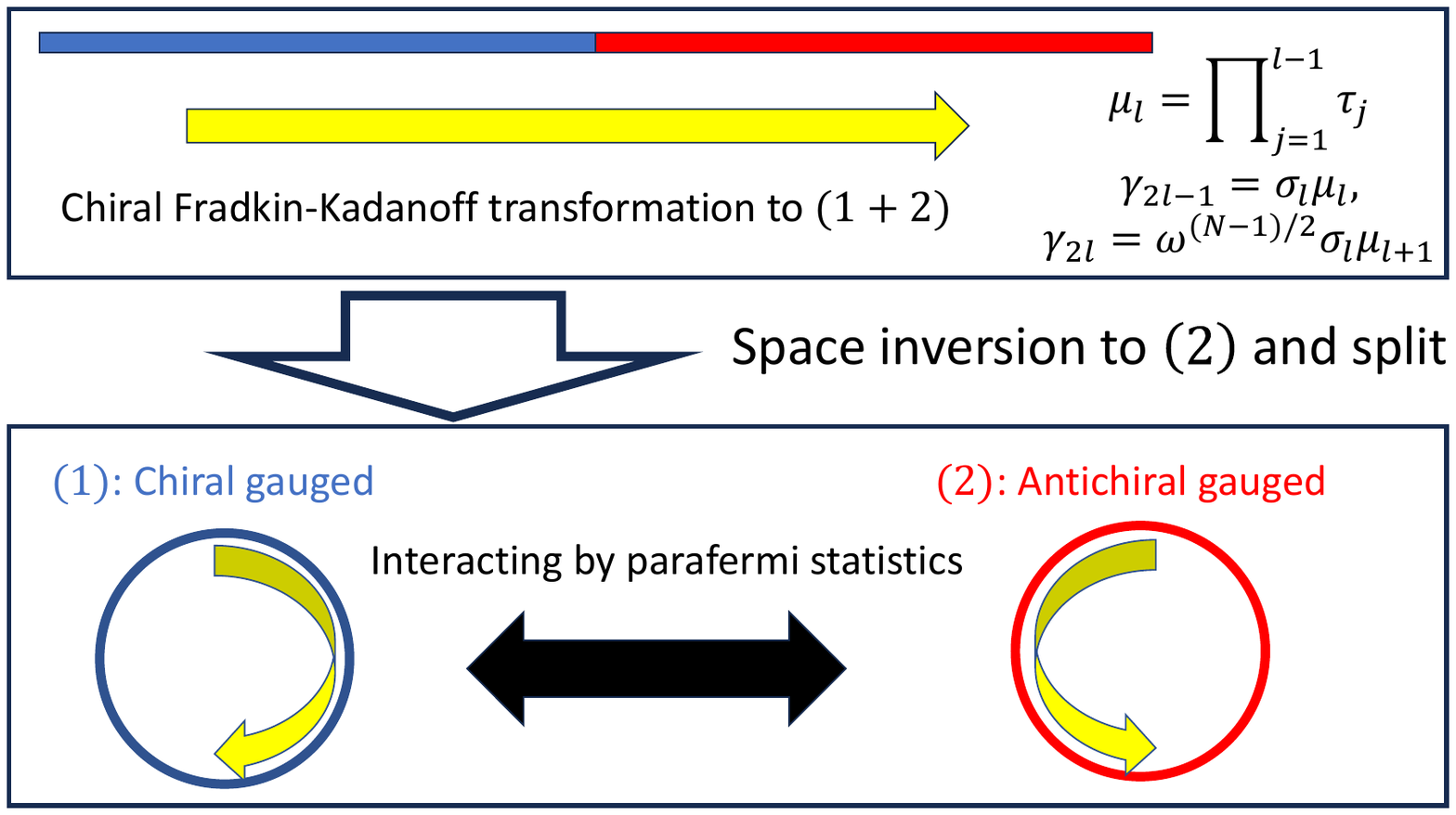}
\caption{The construction of nonchiral gauging is implemented by gauging, inversion, and splitting. By applying Fradkin-Kadanoff transformations to coupled spin chains labelled by $(1)$ and $(2)$ and space inversion operation only to spin chain (2), one can obtain the new composite $Z_{N}$ particle models. It should be noted that the space inversion is not included in a conformal transformations.}
\label{Dirac_type_parefermion}
\end{center}
\end{figure}

The new gauging procedure can be easily implemented to the two copies of identical paraspin chains labeled by (1) and (2), by applying the chiral Fradkin-Kadanoff transformation to the chain (1) and antichiral Fradkin-Kadanoff transformation to the chain (2). More detailed procedures are as follows.
\begin{itemize}
\item{Applying  the chiral Fradkin-Kadanoff transformation to a spin chain $(1+2)$ with the site labelled by $l=1, 2, ...,2L$. After the Fradkin-Kadanoff transformation, one obtains the site $l_{\text{PF}}=1,2,...,4L$.}
\item{Splitting the chain $(1+2)$ to chains $(1)$ and $(2)$ with the identification of each site $l^{(1)}=l_{\text{PF}}$, $l^{(2)}=4L+1-l_{\text{PF}}$ where we have taken the number of site of each chain as $l^{(1)}=1,2,...2L$, $l^{(2)}=1,2,...2L$.}
\end{itemize}
The construction is summarized also in Fig. \ref{Dirac_type_parefermion}. After the new parafermionization, one can implement the following composite particle,
\begin{equation}
\Gamma_{i}=\gamma_{i}^{(1)}\gamma_{i}^{(2)}
\end{equation}
where we have labelled the variabele in the respective chain by $(1)$ and $(2)$ and they play the roles like UV and IR in the main text.
This composite particle satisfies the following relations,
\begin{equation}
\Gamma_{i}\Gamma_{j}=\Gamma_{j}\Gamma_{i}, \ \Gamma_{i}^{N}=1
\end{equation}
Because of these bosonic mutual statistics and $Z_{N}$ property, one can identify the variable $\{ \Gamma \}$ as the lattice analog of integer spin simple current with nonchiral $Z_{N}$ particles.

Because of the nonlocal property of the Fradkin-Kadanoff transformation\cite{Yao:2020dqx}, the coupled parafermion chains are strongly interacting after the transformations even when starting from a local Hamiltonian in the original paraspin representation. This property comes from the parafermi statistics of the chains. As the simplest example, we note the results for the $Z_{3}$ parafermion corresponding to the coupled three state Potts model.
\begin{align}         
Z_{0}=\sum_{i^{(1)},i^{(2)}\in \{ I, \epsilon\}^{2}}|\sum_{p=0}^{2}\chi^{(1)}_{i^{(1)},p}\overline{\chi}^{(2)}_{\overline{i}^{(2)},p}|^{2}, \\
Z_{2/3}=\sum_{i^{(1)},i^{(2)}\in \{ I, \epsilon\}^{2}}|\sum_{p=0}^{2}\chi^{(1)}_{i^{(1)},p+1}\overline{\chi}^{(2)}_{\overline{i}^{(2)},p}|^{2}, \\
Z_{1/3}=\sum_{i^{(1)},i^{(2)}\in \{ I, \epsilon\}^{2}}|\sum_{p=0}^{2}\chi^{(1)}_{i^{(1)},p+2}\overline{\chi}^{(2)}_{\overline{i}^{(2)},p}|^{2},
\end{align}

where the primary fields in the single $Z_{3}$ model are $\{ I, \psi_{1}, \psi_{2}, \epsilon, \sigma_{1}, \sigma_{2} \}$ and $\psi$ is the $Z_{3}$ simple current known as chiral parafermion and the lower index labels the parafermionic parity. Their respective conformal dimensions are $\{ h_{I}=0, h_{\psi_{1}}=2/3, h_{\psi_{2}}=2/3, h_{\epsilon}=2/5, h_{\sigma_{1}}=1/15, h_{\sigma_{2}}=1/15 \}$, and $\epsilon$ satisfies the fusion rule of Fibonacci anyon, $\epsilon \times \epsilon =I+\epsilon$. As in the usual modular covariant, the charge index $0, 1/3, 2/3$ corresponds to the $Z_{3}$ twisted boundary conditions of the coupled chain.

If the two $Z_{N}$ CFTs and the corresponding lattice models are in the same anomaly classifications, the above analysis can apply to the coupled system. This is quite natural from the anomaly classifications, but we stress that the form of modular invariants and the corresponding wavefunctions of TO are nontrivial. We also note the resultant partition functions of other simple CFT, coupled $SU(3)_{1}$ WZW model, 
\begin{align}
Z_{0}=|\chi^{(1)}_{I}\overline{\chi}^{(2)}_{I}+\chi^{(1)}_{J_{1}}\overline{\chi}^{(2)}_{J_{1}}+\chi^{(1)}_{J_{2}}\overline{\chi}^{(2)}_{J_{2}}|^{2}, \\
Z_{0}=|\chi^{(1)}_{J_{1}}\overline{\chi}^{(2)}_{I}+\chi^{(1)}_{J_{2}}\overline{\chi}^{(2)}_{J_{1}}+\chi^{(1)}_{I}\overline{\chi}^{(2)}_{J_{2}}|^{2}, \\
Z_{0}=|\chi^{(1)}_{J_{2}}\overline{\chi}^{(2)}_{I}+\chi^{(1)}_{I}\overline{\chi}^{(2)}_{J_{1}}+\chi^{(1)}_{J_{1}}\overline{\chi}^{(2)}_{J_{2}}|^{2}.
\end{align}
where $J_{1}$ and $J_{2}$ are the $Z_{3}$ simple current with conformal dimension $1/3$. The relationship between the above $SU(3)_{1}\times SU(3)_{1}$ and the coset representation of the Fateev-Zamolochikov parafermion model, or $SU(3)_{1}\times SU(3)_{1}/SU(3)_{2}$, is a future problem. We also note that the partial exchange of the lower index $J_{1}^{(1)}\leftrightarrow J_{2}^{(1)}$ leads to the different topological phase, because the generator of the gauging operations is different. Hence even when starting from the identical spin chain, one can obtain a different phase diagram by introducing perturbations.

Construction of the BCFT for our models is an interesting problem, but one can obtain the corresponding OECFT by applying the method in \cite{Fukusumi:2020irh} and the main text. By introducing the parity and charge for the simple current $J^{(1)}\overline{J}^{(2)}$ and studying the $Z_{N}$ extension of the spherical fusion category corresponding to the original bosonic CFT $(1)+ (2)$ (or UV$\times$IR in the main text) and by taking a particular subalgebra called bulk semionization, one can obtain the algebraic data corresponding to the Cardy's condition.

For this purpose, let us assume the parity structure of $Z_{N}$ extended SFC and denote the anyonic objects (or bulk primary fields) as $\Phi_{i, p^{\text{L}}, p^{\text{R}}}$, $\Phi_{a, p}$, where we introduced $p^{\text{L}}$,  $p^{\text{R}}$ to introduce the parity structure for $J^{(1)}\overline{J}^{(2)}$ and $\overline{J}^{(1)}J^{(2)}$. One may have noticed that there exists no distinction between the left edge and right edge for the sector labeled by $a$ and this is a consequence of entanglement. Then one can obtain the algebraic data of OECFT, or its symmetry topological field theory (SymTFT) as,

\begin{align}
\Psi_{i,p}&= \frac{\sum_{p}\Phi_{i,p+p',-p'}}{N}\\
\Psi_{a,p}&=\Phi_{a,p}
\end{align}
The forms are the same as those by the author's other work \cite{Fukusumi:2024cnl} and the $\{ \Psi\}$ in the main text. Hence, by identifying the respective object as $\{\phi^{(1)}\}\otimes\{ \overline{\phi}^{(2)}\}$ or its complex conjugate, one can obtain the resultant fusion rule of OECFT. 

Here we also note the form of the operators constructing the correlation function of OECFT as,

\begin{align}
&\phi_{i_{\text{tot}},p_{L}}, \\
&\{ \Phi_{a_{\text{tot}},p}\}_{SD}=\sum_{p'}\frac{\phi_{a_{\text{tot}},p'+p}\phi_{a_{\text{tot}}-p'}}{\sqrt{a}}
\end{align}
where SD in the lower index is the doubling trick\cite{PhysRevLett.54.1091,Cardy:1986gw} to bring the operator in the right edge to left edge or the mapping $\{\overline{\phi}^{(1)}\}\otimes\{ \phi^{(2)}\} \rightarrow \{\phi^{(1)}\}\otimes\{ \overline{\phi}^{(2)}\}$.

\section{ Locality versus mock modularity in the construction of wavefunctions ($2$-dimensional gauge theory coupled to matter)}
\label{nonchiral_wavefunction}
In the main text, we have concentrated our attention on the matter part of the CFT when identifying an FQHE as $CFT\times \text{flux}$. However, by using the existing formalism, one can construct the nonchiral correlation function corresponding to FQHEs with nonchiral anyons by introducing the effect of the flux. We expect that this corresponds to the wavefunctions of FQHE in plane geometry and we mainly follow the approach in \cite{Ino:1998by,Milovanovic:1996nj}. To simplify the discussion, let us consider the OECFT coupled with the chiral $U(1)$ flux $CCFT^{(1)}\times \overline{CCFT}^{(2)}\times U(1)$. We assume that each matter CFT has chiral and antichiral $Z_{2}$ simple current $J^{(1)}$ and $\overline{J}^{(2)}$ satisfying the anomaly-free condition $Q_{J^{(1)}\overline{J}^{(2)}}\left(J^{(1)}\overline{J}^{(2)}\right)=0$. Successive to Appendix \ref{Lattice_realization}, the upper index (1) and (2) correspond UV and IR in the main text. We also assume that the $U(1)$ theory produces the Laughlin wavefunction without coupling to the matter and the form of an electric object without matter coupling as $e^{i\sqrt{q}\varphi}$ where $\varphi$ is a bosonic field and $q$ is an integer. 

The $N$ electron wavefunction without coupling to matter is,
\begin{equation}
\langle e^{-N i\sqrt{q}\varphi (\infty)}\prod_{i}^{N}  e^{i\sqrt{q}\varphi (z_{i})}\rangle =\prod_{i<j} \left(z_{i}-z_{j} \right)^{2q}
\end{equation}
where $e^{-N i\sqrt{q}\varphi (\infty)}$ is the so-called background charge or label of edge modes.The quasihole operator without matter is $e^{i\frac{r}{\sqrt{q}}\varphi}$ where $r=0, 1, ...q-1$. The above wavefunction is known as Laughlin's wavefunction\cite{Laughlin:1983fy}.

Then the nonchiral correlation function corresponding to this nonchiral FQHE is,

\begin{equation}
\langle e^{-N i\sqrt{q}\varphi (\infty)}\prod_{i}^{N}  e^{i\sqrt{q}\varphi (z_{i})}J^{(1)}(z_{i})\overline{J}^{(2)}(\overline{z}_{i})\rangle
\end{equation}
for an even $N$ and

\begin{equation}
\langle J^{(1)}(\infty)\overline{J}^{(2)}(\infty)e^{-N i\sqrt{q}\varphi (\infty)}\prod_{i}^{N}  e^{i\sqrt{q}\varphi (z_{i})}J^{(1)}(z_{i})\overline{J}^{(2)}(\overline{z}_{i})\rangle
\end{equation}
for an odd $N$.

One can consider the quasihole operator by considering the total $Z_{2}$ charge. The form is simple,
\begin{equation}
e^{i\frac{r+Q_{J_{\text{tot}}}(\alpha^{(1)}\overline{\alpha}^{(2)})}{\sqrt{q}}\varphi}\phi^{(1)}_{\alpha^{(1)}}\overline{\phi}^{(2)}_{\overline{\alpha}^{(2)}}
\end{equation}
where 
\begin{equation}
\begin{split}
Q_{J_{\text{tot}}}(\alpha^{(1)}\overline{\alpha}^{(2)})&=Q_{J^{(1)}}(\alpha^{(1)})-Q_{\overline{J}^{(2)}}(\overline{\alpha}^{(2)}) \\
&=h_{J^{(1)}}+h_{\alpha^{(1)}}-h_{J^{(1)}\times \alpha^{(1)}} \\
&-\overline{h}_{\overline{J}^{(2)}}+\overline{h}_{\overline{\alpha}^{(2)}}-\overline{h}_{\overline{J}^{(2)}\times \overline{\alpha}^{(2)}}
\end{split}
\end{equation}
and $\phi$ and $\overline{\phi}$ are the corresponding chiral and antichiral fields in the matter CFTs.

Because of the existence of the twisted sector of the matter CFTs, the $U(1)$ part obtains a half flux quantum. Moreover, by applying the modular $S$ transformation to the corresponding characters, one can see the appearance of a mock modular form\cite{Ino:1998by,Milovanovic:1996nj} (whereas they have not used this mathematical terminology for the characters).

One can apply the above analysis to general $Z_{N}$ models only by replacing some symbols in the existing literature (see review \cite{Hansson_2017}, and the general construction in \cite{Schoutens:2015uia}). This can include the case where the matter CFTs are $Z_{N}$ anomalous, but coupling this anomalous matter to flux, one can obtain the modified anomaly-free theory by modifying $J_{\text{tot}}$ to $J_{\text{tot}}\times (\text{flux})$. Hence, the total partition function is still in the form Eq. \eqref{universal_partition_function} with the modified total charge zero. Moreover, contrary to the case with anomaly-free matter, the theory does not contain mock-modular forms. However, it has been pointed out that the resultant partition function is not in the Moore-Seiberg data\cite{Cappelli:1996np}. Hence, there exists difficulty to obtain bulk topological degeneracies from the existing framework in this case because of the close relationship between Moore-Seiberg data and bulk topological orders. Related problems have been summarized in the author's other works \cite{Fukusumi:2022xxe,Fukusumi_2022,Fukusumi_2022_c}. Hence, there exists the following exchange relation
\begin{equation}
\begin{split}
\{\text{Anomaly free matter $\Rightarrow$ mock modular form}\} \\
\leftrightarrow \{\text{Anomalous matter $\Rightarrow$ Non Moore-Seiberg}\} 
\end{split}
\end{equation}
where $\Rightarrow$ represents coupling to the flux.

Before going to the next section we note an interesting property of $Z_{2}$ and $Z_{4}$ symmetric models.
The anomaly-free condition for $CCFT^{(1)}\times \overline{CCFT}^{(2)}$ and $CCFT^{(1)}\times CCFT^{(2)}$ are the same for $Z_{2}$ and $Z_{4}$ cases. This property of $Z_{2}$ and $Z_{4}$ might be useful for further studies. In the $Z_{2}$ case, this seems to correspond to the particle-hole conjugation\cite{Son:2015xqa,Barkeshli:2015afa,Wan2015StripedQH,Yang_2017}. For example, this conjugation exchanges Pfaffian and anti-Pfaffian wavefunctions with each other\cite{Levin_2007,Lee_2007,PhysRevLett.117.096802}.

\section{Conjecture on the boundary states corresponding to the RG and charged domain walls}
\label{conjecture_BCFT}
In this section, based on the arguments in the main text, we conjecture the form of boundary states corresponding to both RG and charged domain walls. However, the determination of the RG domain wall requires more specific data from the models, such as boundary $g$-function\cite{Affleck:1991tk} and related transmission or reflection coefficient\cite{Quella:2006de,Kimura:2014hva,Kimura:2015nka}.

By applying the construction of $Z_{N}$ gauged boundary states to the coupled model, one can obtain the following forms,

\begin{align} 
|i_{\text{tot}}\rangle\rangle_{Z_{N}}^{(l)}&=\sum_{p} \omega^{lp} |i_{\text{tot}},p\rangle\rangle ,\\
|a_{\text{tot}}\rangle\rangle_{Z_{N}}^{(l)}&=\sum_{p}  \frac{\omega^{lp}}{\sqrt{N}}|a_{\text{tot}},p\rangle\rangle
\end{align}
where their Cardy condition is,
\begin{equation}
\langle \langle \alpha_{\text{tot}}| e^{i\pi \tau H_{\text{CFT,tot}}}|\alpha'_{\text{tot}}\rangle\rangle=\sum_{\gamma_{\text{tot}}} n^{\gamma_{\text{tot}}}_{\alpha_{\text{tot}} \alpha'_{\text{tot}}}\Xi_{\gamma_{\text{tot}}}(-1/\tau)
\end{equation}
We remind that the above characters are determined by the original product theories and this implies that the problem of the RG domain wall now becomes more tractable.
We also note the corresponding forms for the $D$-type theories by applying the methods in \cite{Runkel:2020zgg,Smith:2021luc,Weizmann,Fukusumi:2021zme},
\begin{align} 
|i_{\text{tot},D}\rangle\rangle_{Z_{N}}^{(l)}&=\sum_{p} \frac{\omega^{lp}}{\sqrt{N}} |i_{\text{tot}},p\rangle\rangle ,\\
|a_{\text{tot},D}\rangle\rangle_{Z_{N}}^{(l)}&=\sum_{p}  \omega^{lp}|a_{\text{tot}},p\rangle\rangle
\end{align}
These boundary states will be useful in formulating and studying RG rigorously and phenomenologically because one can map a problem of RG (which requires a lot of analytical calculations combined with numerical simulations) to the representation theory of conformal interfaces determined algebraically. Moreover, one can calculate the respective quantities by representing the above labels as $i_{\text{tot}}$ and $a_{\text{tot}}$ tensor products of underlying theories.

\section{Bulk and boundary RG flow: Application of Graham-Watts method}

To analyze the stability of edge modes of $1+1$ dimensional system, the bulk and boundary RG analysis is significant. If one assumes the two systems labeled by UV and IR are connected under massless RG flow corresponding to RG domain wall $D_{\text{RG}}$, the following transformation for boundary states should provide a canonical formulation of bulk and boundary RG flows,
\begin{equation}
 D_{\text{RG}}| \alpha_{\text{UV}}\rangle\rangle= \sum_{\alpha_{\text{IR}}}D_{\text{RG},\alpha_{\text{UV}}}^{(\text{def})\alpha_{\text{IR}}}|\gamma_{\text{IR}}\rangle\rangle
\end{equation}
where $\alpha$ and $\gamma$ are the primary states labeling the UV and IR Cardy states 
and the matrix $D_{\text{RG}}^{\text{def}}$ is the matrix representation of the RG domain wall (We demonstrate the calculations elsewhere). The most fundamental point of this equation is the righthand side can become nonCardy states (such as Graham-Watts states\cite{Graham:2003nc,Fukusumi:2020irh}) because the boundary $g$-value can show nondecreasing under bulk and boundary RG flow\cite{Green:2007wr}. Similar application of the charged domain wall to boundary states is an interesting future problem. We also note that the original motivation introducing RG domain wall has a close connection to the bulk and boundary RG flow\cite{Brunner:2007ur}.

\bibliographystyle{ytphys}
\bibliography{smeared_Cardy_v5}

\end{document}